%% file: manuscript.tex
\documentclass[]{aastex631}

\newcommand{\jwst}{JWST}

\newcommand{\htwo}{H$_2$}
\newcommand{\hi}{\ion{H}{1}}

\newcommand{\feii}{[\ion{Fe}{2}]}
\newcommand{\neii}{[\ion{Ne}{2}]}

\newcommand{\asec}{$^{\prime\prime}$}
\newcommand{\amin}{$^{\prime}$}
\newcommand{\gradi}{$^{\circ}$}

\newcommand{\kms}{km\,s$^{-1}$}

\newcommand{\um}{$\mu$m}

\newcommand{\msun}{M$_{\odot}$}

\begin{document}

\title{PROJECT-J: JWST observations of HH46~IRS and its outflow. Overview and first results \footnote{}}

\correspondingauthor{Brunella Nisini}
\email{brunella.nisini@inaf.it}

\author[0000-0002-9190-0113]{Brunella Nisini}
\affiliation{ INAF - Osservatorio Astronomico di Roma, 
Via di Frascati 33, 00078 Monte Porzio Catone, Italy}

\author[0000-0002-1860-2304]{Maria Gabriela Navarro}
\affiliation{ INAF - Osservatorio Astronomico di Roma, 
Via di Frascati 33, 00078 Monte Porzio Catone, Italy}

\author[0000-0002-7035-8513]{Teresa Giannini}
\affiliation{ INAF - Osservatorio Astronomico di Roma, 
Via di Frascati 33, 00078 Monte Porzio Catone, Italy}

\author[0000-0002-0666-3847]{Simone Antoniucci}
\affiliation{ INAF - Osservatorio Astronomico di Roma, 
Via di Frascati 33, 00078 Monte Porzio Catone, Italy}

\author[0000-0001-6872-2358]{Patrick, J. Kavanagh}
\affiliation{Department of Experimental Physics, Maynooth University, 
Maynooth, Co Kildare, Ireland}

\author[0000-0002-5380-549X]{Patrick Hartigan}
\affiliation{Physics and Astronomy Dept., Rice University, 
6100 S. Main, Houston, TX 77005-1892, USA}

\author[0000-0001-5776-9476]{Francesca Bacciotti}
\affiliation{INAF - Osservatorio Astrofisico di Arcetri, 
Largo E. Fermi 5, I-50125 Firenze, Italy}

\author[0000-0001-8876-6614]{Alessio Caratti o Garatti}
\affiliation{INAF - Osservatorio Astronomico di Capodimonte, via Moiariello 16, 80131 Napoli, Italy}

\author[0000-0002-6296-8960]{Alberto Noriega-Crespo}
\affiliation{Space Telescope Science Institute, 3700 San Martin Drive, Baltimore, MD, 21218, USA}

\author[0000-0001-7591-1907]{Ewine F. van Dishoeck}
\affiliation{Leiden Observatory, Leiden University, PO Box 9513, NL 2300, RA Leiden, The Netherlands}

\author[0000-0002-3741-9353]{Emma T. Whelan}
\affiliation{Department of Experimental Physics, Maynooth University, Maynooth Co. Kildare, Ireland}

\author[0000-0001-5653-7817]{Hector G. Arce}
\affiliation{Department of Astronomy, Yale University, New Haven, CT 06511, USA}

\author[0000-0002-1593-3693]{Sylvie Cabrit}
\affiliation{LERMA, Observatoire de Paris-PSL, Sorbonne Université, CNRS,  F-75014 Paris, France}
\affiliation{IPAG, Observatoire de Grenoble, Université Grenoble-Alpes, France}

\author[0000-0002-2210-202X]{Deirdre Coffey}
\affiliation{University College Dublin, School of Physics, Belfield, Dublin 4, Ireland}

\author[0000-0001-6156-0034]{Davide Fedele}
\affiliation{INAF - Osservatorio Astrofisico di Arcetri, Largo E. Fermi 5, I-50125 Firenze, Italy}

\author[0000-0001-6496-0252]{Jochen Eisl\"offel}
\affiliation{Thüringer Landessternwarte, Sternwarte 5, D-07778 Tautenburg, Germany}

\author[0000-0002-9122-491X]{Maria Elisabetta Palumbo}
\affiliation{INAF - Osservatorio Astrofisico di Catania, via Santa Sofia 78, 95123, Catania, Italy}

\author[0000-0003-2733-5372]{Linda Podio}
\affiliation{INAF - Osservatorio Astrofisico di Arcetri, Largo E. Fermi 5, I-50125 Firenze, Italy}

\author[0000-0002-2110-1068]{Tom P. Ray}
\affiliation{Dublin Institute for Advanced Studies, 31 Fitzwilliam Place, D02 XF86, Dublin, Ireland}

\author{Megan Schultze}
\affiliation{Physics and Astronomy Dept., Rice University, 
6100 S. Main, Houston, TX 77005-1892, USA}

\author[0000-0001-6926-1434]{Riccardo G. Urso}
\affiliation{INAF - Osservatorio Astrofisico di Catania, via Santa Sofia 78, 95123, Catania, Italy}

\author[0000-0001-8657-095X]{Juan M. Alcal\'a}
\affiliation{INAF - Osservatorio Astronomico di Capodimonte, via Moiariello 16, 80131 Napoli, Italy}

\author{Manuel A. Bautista}
\affiliation{Department of Physics, Western Michigan University, Kalamazoo, MI 49008, USA}

\author[0000-0003-1514-3074]{Claudio Codella}
\affiliation{INAF - Osservatorio Astrofisico di Arcetri, Largo E. Fermi 5, I-50125 Firenze, Italy}

\author[0000-0002-8963-8056]{Thomas P. Greene}
\affiliation{NASA Ames Research Center, MS 245-6, Moffett Field, CA 94035, USA}

\author[0000-0003-3562-262X]{Carlo F. Manara}
\affiliation{European Southern Observatory, Karl-Schwarzschild-Strasse 2, 85748 Garching, Germany}

\begin{abstract}

We present the first results of the JWST program PROJECT-J (PROtostellar JEts Cradle Tested with \jwst ), designed to study the Class I source HH46~IRS and its outflow through NIRSpec and MIRI spectroscopy (1.66 to 28 \um). The data provide line-images ($\sim$ 6\farcs 6 in length with NIRSpec, and up to $\sim$ 20$^{\prime\prime}$ with MIRI) revealing unprecedented details within the jet, the molecular outflow and the cavity. We detect, for the first time, the red-shifted jet within $\sim$ 90 au from the source. Dozens of shock-excited forbidden lines are observed, including highly ionized species such as [\ion{Ne}{3}] 15.5 \um , suggesting that the gas is excited by high velocity ($>$ 80 \kms) shocks in a relatively high density medium. Images of \htwo\ lines at different excitations outline a complex molecular flow, where a bright cavity, molecular shells, and a jet-driven bow-shock interact with and are shaped by the ambient conditions. Additional NIRCam 2\um\ images resolve the HH46~IRS $\sim$ 110 au binary system and suggest that the large asymmetries observed between the jet and the \htwo\ wide angle emission could be due to two separate outflows being driven by the two sources. The spectra of the unresolved binary show deep ice bands and plenty of gaseous lines in absorption, likely originating in a cold envelope or disk. In conclusion, \jwst\ has unraveled for the first time  the origin of the HH46~IRS complex outflow demonstrating its capability to investigate embedded regions around young stars, which remain elusive even at near-IR wavelengths.

\end{abstract}

\section{Introduction} \label{sec:intro}

The study of protostars and their planet-forming disks is inseparable from understanding the role and properties of their associated outflows.
Protostellar systems include several components (i.e.\ the stellar object, a compact and massive accretion disk, and a dusty envelope) but the most prominent phenomenon is mass ejection in the form of powerful jets and winds. Their strong line emission extends up to parsec scales from the protostar, and dominates the spectrum at almost all wavelengths \citep{bally2016,frank2014}.
It is now widely accepted that jets are magneto-centrifugally launched from an inner star-disk interaction region \citep[e.g.][]{pelletier1992}, although the details of the mechanism are still poorly constrained, especially during the early phases of stellar evolution. According to MHD disk-wind models \citep[see e.g,][]{ferreira1997,bai2016}, matter is extracted at different spatial radii from the disk surface. These ejections remove disk mass and angular momentum, and thus they have a fundamental role in driving accretion and determining the time-scale of disk dispersal. The high-velocity (200-400 \kms) collimated jets are launched from the very inner regions of the disk (on scales of a few 0.1 au), however mass loss in the form of less collimated and slow flows (10-30 \kms) might involve a large portion of the disk (up to several tens of au), thus affecting the disk physics and influencing the formation of planetary systems \citep[e.g.][]{pascucci2023}. Alternative models, such as the X-wind model, assume that the outflow originates from a small region at the interface between the inner edge of the disk and the stellar magnetosphere. They also predict the formation of a collimated high-velocity outflow surrounded by a wide-angle wind \citep{shang2020}. 
The manner in which jets and winds are launched and propagate also influences their feedback on the ambient medium and envelope dispersal during the early stages with the typical creation of large cavities in the infalling envelopes and the entrainment of the cold ambient gas \citep[e.g][]{Rabenanahary2022,shang2023} which is observed in the sub-mm by, e.g, ALMA \citep[e.g.][]{arce2013,devalon2022}.

To get insights into these different mechanisms one needs to peer into regions within a few hundreds of au from the central star, where the outflows preserve the pristine information about their velocity, collimation mechanism and connection with accretion events. However, during the critical stage of protostellar vigorous accretion, the innermost regions of the jet are still deeply embedded in their dusty envelope making their investigation at high angular resolution challenging. Additionally, when studying outflows in Young Stellar Objects (YSOs) of different ages, one should bear in mind that the physical conditions of both the ejections and the environment through which the flow travels vary considerably during evolution. Consequently, the choice of observational tracers should be carefully tailored depending on the source.
In particular, while jets in more evolved and less extincted Classical T Tauri stars (CTTs, Class II sources) have mostly been studied with optical forbidden lines, the origin of outflows from embedded Class 0 and Class I protostars (ages 1-10$\times$10$^4$ yr) needs to be investigated in the IR and sub-mm domain, as this is the spectral range where most of the outflowing gas is emitting. 
For the youngest and deeply embedded Class 0 sources, high-resolution sub-mm observations with ALMA are getting insights into the mostly molecular and cold jets \citep[e.g.][]{lee2020,podio2021}. However, the warmest gas component expected to be excited, either by shocks or by the action of UV photons, along protostellar flows and associated cavities is best traced through IR atomic and molecular lines \citep{frank2014}, as highlighted by observations with previous space facilities such as the Infrared Space Observatory (ISO), and the Spitzer and Herschel satellites \citep[e.g.][]{nisini2003,giannini2011,watson2016,nisini2016}. It is however only thanks to the superb resolution and sensitivity of the James Webb Space Telescope (JWST) that it is now possible to investigate the launching mechanism of jets/outflows in the earliest stages, and their feedback on the ambient medium, with the same level of accuracy as currently achieved only in the optical for the more evolved CTT stars.  

JWST observations of Class 0 sources are indeed now giving unprecedented details on the morphology and composition of their jets, wide angle molecular flows and cavities \citep[e.g.][]{federman2023}, although in the youngest of these protostars their highly embedded innermost region can be hardly traced even at Mid-IR (MIR) wavelengths \citep{ray2023}. On the other hand, JWST is best suited for the study of the intermediate age Class I objects (age $\sim 10^5$ yr), which are still actively accreting from a massive disk, but 
their parental envelope, within which they are still embedded, can be easily penetrated at mid-IR wavelengths. 
This in fact makes the Class I sources the only sources during the young star's evolution where all the actors involved, namely the fast axial jet, the wide-angle slow wind, the accreting protostar and the inner disk region, can be all efficiently studied with JWST. 

With this in mind, we present here the PROtostellar JEts Cradle Tested with \jwst\ (PROJECT-J) project, a Cycle 1 GO program aiming at studying the Class I protostar HH46~IRS and its outflow system through Integral Field Spectroscopy (IFU) observations from 1.7 to 28 \um\, with NIRSpec and MIRI. The main aim of the project is to disentangle among the proposed mechanisms at the origin of the outflow and the associated cavity. This will be achieved providing a unified view of all the outflow components pertaining to this paradigmatic Class I system, and connecting them with the accretion and stellar properties of the central object, which can now be characterised in the MIR. This paper provides a general overview of the program, describes the reduction and quality of the acquired data, and highlights its main outcomes. A more detailed analysis of the different scientific topics will be presented in specific papers currently in preparation. 
The paper is structured as follows: Sect. 1.1 gives an overview of the HH46 source and its outflow; Sect. 2 presents the observations and data reduction for the MIRI (Sect. 2.1) and NIRSPEC (Sect. 2.2) dataset, with a focus on the process adopted to produce spectra (Sect. 2.3) and line emission maps (Sect. 2.4); Sect. 3 illustrates the results concerning the outflow, i.e. the line maps of the atomic jet (Sect. 3.1), and the molecular outflow and cavity (Sect. 3.2); Sect. 4 presents the spectrum of the protostar; finally, Sect. 5 presents the main conclusions of the performed study.

\subsection{HH46~IRS and its outflow}

HH46~IRS (2MASSJ08254384-5100326) is a low mass Class I source, (M $\sim 1.2 $\msun, $L_{bol} <$ 15 L$_\sun$, \cite{antoniucci2008}), located in a Bok globule at the edge of the Gum Nebula ($d$=450 pc). Hubble Space Telescope (HST) observations suggest that the source is actually an embedded binary system with a projected separation of 0\farcs 26 ($\sim$ 120 au) \citep{reipurth2000}. The source drives a parsec scale chain of Herbig Haro (HH) objects known as the HH46/47 outflow system, which has been studied at both optical \citep[e.g.][]{eisloffelpm1994,heathcote1996,hartigan2011} and infrared wavelengths \citep[e.g.][]{eisloffelh21994,reipurth2000,noriega2004,erkal2021}. The optically bright HH 46/47 jet is part
of the northeast flow and terminates in the HH 47A bow shock at about 1\farcm 3 from the source. Multi-epoch HST observations of this jet have shown that the jet moves at $\sim$ 300 \kms\, and it is inclined by about 37\gradi\ with respect to the plane of the sky \citep{hartigan2005}. The jet presents a large wiggling morphology, believed to originate from precession around a tertiary non-resolved source \citep{reipurth2000}, which also causes a change with time of the jet orientation angle of about 15\gradi\ \citep{hartigan2005}. 
A fainter counter-jet in the red-shifted southwestern lobe \citep{eisloffelh21994,velusamy2007} extends toward the bow shock HH 47C. 
Recent deep \feii\, HST images of the outflow have revealed unprecedented details of the inner region of the jet, showing that the blue-shifted jet is formed by several collimated emission knots displaced along an arc-shaped pattern that follows the general curved morphology at larger scales \citep{erkal2021}. Even in the near-IR, the red-shifted jet remains obscured close to the source and only emerges at a distance of about 5\asec\ appearing highly misaligned with respect to the blue-shifted jet. This red-shifted jet has also been detected at mid- and far-IR wavelengths although at low spatial resolution \citep{velusamy2007,nisini2015}. The large contamination from the central source however prevents tracing the counter-jet close to the source in these images.
The large-scale HH46~IRS outflow has been mapped at sub-millimeter wavelengths in CO 6-7 and [\ion{C}{1}] 2-1 by \cite{vankempen2009}, who suggested that the excitation of these lines could be due to ultraviolet photons originating in the jet shocks.

Recently, the HH46~IRS outflow has been studied via high angular resolution ALMA observations of CO 1-0 \citep{arce2013,zhang2016} and CO 2-1 \citep{zhang2019}. 
These observations have shown the presence of nested shell-like structures of cold molecular gas displaying a regular velocity pattern, such that the higher velocity structures extend further from the source. It has been suggested that such a shell-like outflow is due to the entrainment of ambient gas by a series of outbursts from an intermittent wide-angle wind \citep{zhang2019}, or, alternatively, by a pulsed narrow jet \citep{Rabenanahary2022}. 
The unified picture connecting the HH46~IRS outflow and jet, that \jwst\ is for the first time able to provide, will clarify the different scenarios for the origin of the complex mass ejection mechanism in this source. 

\section{Observations and Data reduction} \label{sec:red}

\begin{figure}[ht]
    \centering
    \includegraphics[width=0.5\textwidth,keepaspectratio]{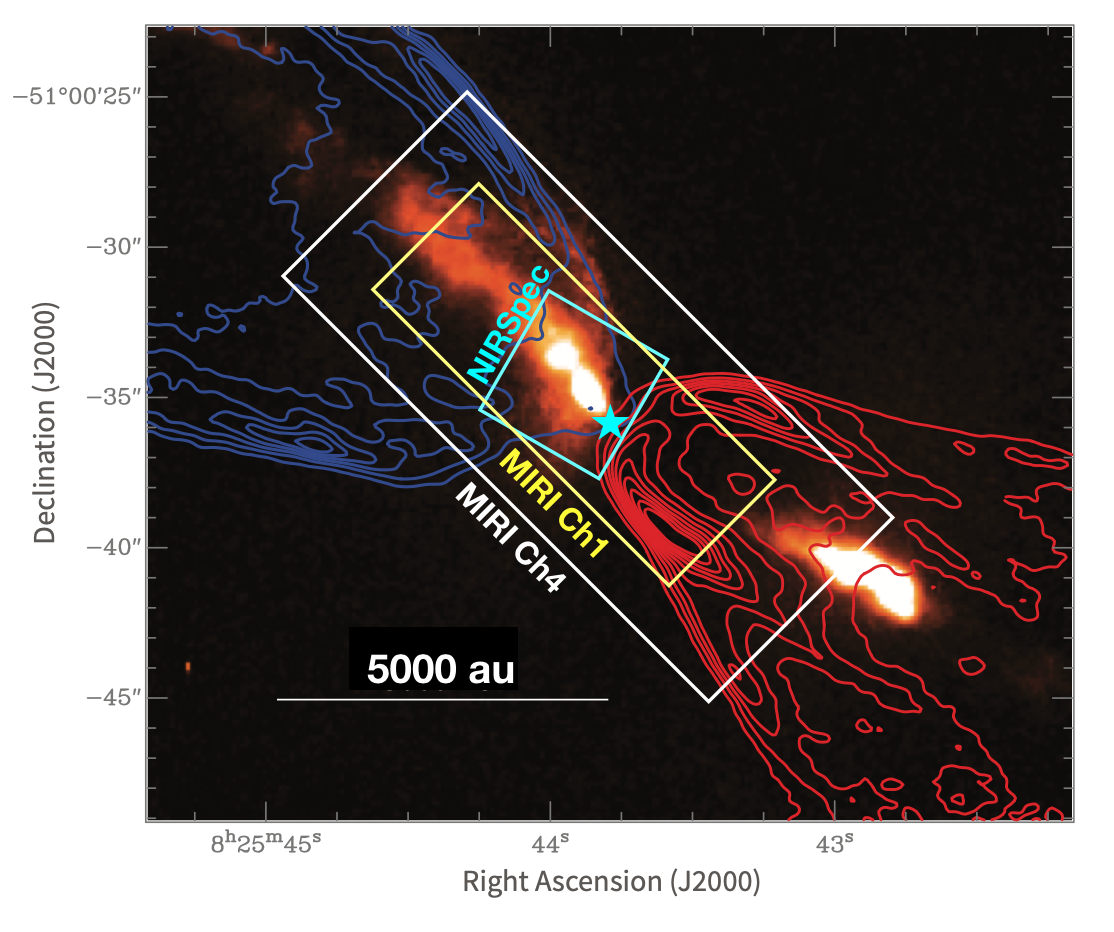}
    \caption{HST WFC3 continuum-subtracted \feii\ 1.64\um\ image \citep{erkal2021} of the HH46 jet, with overlaid contours of the ALMA CO 1-0 red-shifted (red contours) and blue-shifted (blue contours) emission from \cite{arce2013}. The 
    cyan box indicates the region covered by the NIRSpec mosaic while for MIRI the smallest and the largest FOVs covered by Channel 1 and 4 are indicated by the yellow and white boxes, respectively.   }
    \label{fig:FOV}
\end{figure}

We mapped the HH46 IRS source and its outflows as part of the \jwst\ Cycle 1 program PROJECT-J (ID 1706, P.I. B. Nisini) with the instruments NIRSpec~IFU \citep{jakobsen2022,boker2022} and MIRI~MRS \citep{rieke2015,wright2023}.
Fig. \ref{fig:FOV} shows the regions mapped with the two instruments, which include the central source, the jet and counter-jet, and the wide-angle cavity traced by ALMA observations. 
The data were taken on February 16 and 19 2023 with NIRSpec and on March 3 2023 with MIRI. Details on the adopted settings and data reduction procedures are given in the next sub-sections.

\subsection{MIRI} \label{sec:red:MIRI}

The MIRI MRS mode adopted for the observations uses four
integral field units (IFUs) referred to as channels, to continuously 
cover the spectral range between 4.9 and 27.9 \um . Each channel is divided into three sub-bands named SHORT, MEDIUM and LONG.
The spatial sampling of the four channels is 0\farcs196, 0\farcs196, 0\farcs245 and 0\farcs273 per pixel, therefore each channel has a different Field of View (FoV) on the sky that varies between 3\farcs2 $\times$ 3\farcs7 (Channel 1) and  6\farcs6 $\times$ 7\farcs7 (Channel 4).  Fig. \ref{fig:FOV} roughly indicates the smallest (6\asec $\times$15\asec) and largest ($\sim$ 11\asec $\times$20\asec) region covered by our 4$\times$2 map. 
The PA range of the observations was constrained in order to ensure the alignment of the mosaic with the outflow, allowing at the same time some flexibility for scheduling. The final PA is 47.5$^\circ$ NE. 
The MIRI spectral resolution varies between $\sim$ 1500 (V$\sim$ 200\kms , Channel 4) and 3500 (V$\sim$ 85\kms , Channel 1).

The observations were taken in FAST1 readout mode with 19 groups of 3 integrations each, for a total of 655\,s in each exposure, adopting a four-point dither pattern optimized for extended sources. The total exposure time for the full map was 4.25 hrs. 
Simultaneous off-source MIRI imaging was conducted in the F560W filter and used for astrometric registration. A background field located in a clean region selected from inspection of Spitzer images was observed for overall background subtraction. 

The data were reprocessed using the \jwst\ pipeline version 1.11.1 \citep{bushouse2023} and corresponding Calibration Reference Data System (CRDS) context jwst$\_$1094.pmap. This version of the pipeline corrects the decrease in count rates at the MIRI-MRS longest-wavelength channels reported in May 2023\footnote{see \url{https://www.stsci.edu/contents/news/jwst/2023/miri-mrs-reduced-count-rate-update}}. 

The individual raw images were first processed for detector-level corrections providing the calwebb$\_$detector1 products (stage1). 
Then, astrometric correction, flat-fielding and flux calibrations were applied using the Calwebb$\_$spec2 module (stage2). The individual
stage2 images were then resampled and coadded onto a final data-cube through the Calwebb$\_$spec3 processing (stage3). In addition, the following custom steps were applied to increase the quality of the final data-cube.
A background subtraction was performed to the rate files, after the calwebb$\_$detector1 stage. For that, we created a master background using the background observation rate files for each detector. 
The accuracy of MIRI pointing in embedded regions is affected by the paucity of suitable guide stars in the FOV.  To improve the astrometric correction, we used HST + Gaia sources identified in the MIRI simultaneous imaging field, deriving a pointing adjustment of 0\farcs 133 in R.A. and 0\farcs 089 in DEC. 
This correction is still subject to uncertainties in the relative astrometry between the imager and the MRS, which is expected to be $<$ 0\farcs 1 \citep{patapis2023}.

A significant number of warm spurious pixels are left after the calibration step 2 process, which are not removed by the stage 3 outlier$\_$detection step. We identify them in individual exposures at the end of stage2, using the LaCosmic algorithm developed to remove cosmic rays hits from CCD images \citep{vandokkum2001}. We then built a warm pixel mask for each detector, that was added to the Data Quality (DQ) extension of the cal files after calwebb$\_$spec2. 

In the calwebb$\_$spec2 stage we applied the residual$\_$fringe step that corrects for the fringes due to the Fabry-Perot interference between the reflective layers of the detectors, left from the first fringe flat correction. This correction is particularly 
crucial for removing the residual fringe pattern from point sources, where, due to the fact that the MIRI pupil is non-uniformly illuminated, the fringe depth and phase significantly change as a function of the portion of the
point spread function (PSF) that is sampled by the detector pixels \citep{Argyriou2023}. An additional fringe correction is also applied to the final spectrum after the spectral extraction (Kavanagh et al. in prep) .

\subsection{NIRSpec} \label{sec:red:nirspec}

We performed a NIRSpec 4$\times$2 map covering a region of about 6\asec$\times$6\asec\ centred at 08$^h$25$^m$43.98$^s$, $-$51\gradi 00\amin 34.57\asec\ with a pixel scale of 0\farcs 1/pixel. The PA of the mosaic is 71.2\gradi\ NE.  We observed with the NIRSpec gratings/filters G235H/F170LP and G395H/290LP, with a continuous coverage wavelength range between 1.66 and 5.27 \um\, providing a spectral resolution $R\sim 2700$ (i.e. $\sim$ 110 \kms ). 
The observations were taken with an NRSIRS2RAPID readout pattern
with 5 groups of 4 integrations each, using a four-point dither pattern. This implies an exposure time of 1400 s per grating in a single observation. A leakage exposure to correct for failed shutters of the NIRSpec Micro-Shutter Array (MSA) was applied. To reach the desired SNR while avoiding exceeding the data volume in a single visit, we repeated the same observing block twice. The total exposure time in each visit was 2.9 hrs.

The data were processed using the \jwst\ pipeline version 1.11.1 with the corresponding CRDS context file jwst$\_$1094.pmap. Since calwebb$\_$detector1 stage was stable, we ran the pipeline from the rates files. The calwebb$\_$spec2 and 3 are similar to those of MIRI. 
We use the default parameters, carefully inspecting the intermediate products for each step. In particular, we find that the count-rate images (i.e., rate
files) produced after the first stage presented significant vertical patterns associated with correlated noise, as discussed in other works using the NIRSpec~IFU \citep[e.g.][]{perna2023}. To correct for this vertical noise, we obtained the median of each column and subtracted them. This procedure was performed in each rate file before running the calwebb$\_$spec2 module.

Due to the lack of guide stars in the FoV, the final pointing accuracy was $\sim$ 0\farcs2, as estimated by comparing the position of the HH46 IRS source in the G395H grating frames at $\lambda > 4\mu$m, with the corresponding position in the Channel 1 of MIRI. We therefore adjusted the NIRSpec WCS solution taking the MIRI source position as a reference. The pointing adjustment resulted in 0\farcs 114116 in R.A. and 0\farcs 21312 in DEC. We evaluated that no appreciable differential rotation between the NIRSpec and MIRI is present on the basis of the comparison of the overall outflow and nebula morphology at the same wavelengths.  

The stage 3 outlier$\_$detection step leaves a significant number of warm pixels. They have a particular pattern making LaCosmic not efficient in finding them, therefore we applied a different procedure to identify and remove them. In the cal files, we estimated the mean in regions of 5$\times$7 pixels and labelled those deviating from the mean by more than 20$\sigma$ as warm pixels. Remaining outliers in the cube were removed masking the unreliable flat field DQ as DO NOT USE. We then updated the DQ extension of the cal files adding these new flagged pixels. This procedure was carried out for each cal file, allowing the construction of a master warm pixel mask, that indicates the pixels that should not be used in data reduction. 

\subsection{Spectra extraction and flux calibration}\label{sec:extraction}

In order to check the internal consistency of the relative flux calibration between NIRSpec and MIRI sub-bands, we extracted a 1D spectrum at the HH46~IRS position adopting an aperture proportional to the diffraction-limited beam size (1.22$\lambda$/D) so that the aperture increases with wavelength.  
The source appears as point-like in both NIRSpec and MIRI images. Thus the binary detected by \cite{reipurth2000} with a separation of 0\farcs26 remains unresolved within the PSF of these instruments. We considered an aperture of 4 times the beam-size in order to retrieve any flux extending beyond the diffraction-limited beam. Any larger aperture provides differences in the extracted spectrum with less than a 1\% difference.
 
For MIRI, fluxes in the different sub-bands are consistent within 3\% with the exception of Channel 3 MEDIUM, which has a flux 10\% higher with respect to the adjacent sub-bands (see Appendix A). This effect is therefore not dependent on the aperture choice but appears as a pipeline issue. We eventually matched the flux between channels by the ratio of median fluxes in the overlapping wavelengths by applying scale factors starting from the shortest wavelength. 
Fig. \ref{fig:calibration} of Appendix A shows also the Spitzer IRS spectrum of HH46~IRS from \cite{noriega2004}, and other literature mid-IR photometric points for the source, that allows us to evaluate the global flux calibrations. The agreement between the Spitzer and the JWST spectrum decreases with wavelength, with the flux in the Spitzer spectrum at 22 \um\, about 50\% lower than the corresponding MIRI flux. The same trend is observed also with Spitzer, WISE and AKARI photometry. Such a large discrepancy cannot be due to calibration issues. 
Similar differences have been seen with MIRI-MRS in other young sources \citep[see e.g.][]{grant2023,tabone2023,gasman2023}. 
A possible origin for these discrepancies can reside in the intrinsic variability of the source. A certain degree of variability is shown by the mid-IR photometry taken with different facilities, although it is at much lower level. 
Alternatively, it might be that the MIRI fluxes are overestimated using our non-optimal extraction that does not take into consideration the complex PSF of the instrument. Finally, the difference could be also due to a different extraction procedure and background subtraction in the long-slit Spitzer spectrum. This issue will be further investigated in a future work. 

For NIRSpec, the flux within the two gratings matches better than 5\% (lower panel of Fig.14 of Appendix A). There is however about a factor of 1.4 of discrepancy in the overlapping regions between NIRSpec and MIRI (see Appendix A). We rescaled the NIRSpec spectrum to the MIRI one, on the basis that in this way we also get a good agreement with the overall flux density with the Spitzer IRAC photometry at 3.6, 4.5 and 5.8 \um\ (see bottom panel of \ref{fig:calibration}). 

Overall, after rescaling the NIRSpec spectrum, we conclude that the relative calibration between NIRSpec and MIRI and among the different spectral segments is better than $\sim$ 5\% up to $\sim$ 20 \um , while it can be up to 10\% at longer wavelengths. The absolute calibration has however an uncertainty of up to 40\% in the on-source extracted spectrum.

\subsection{Maps of continuum-subtracted line emission}

One of the principal aims of PROJECT-J is to provide continuum-subtracted line images of the HH46 outflow in order to study it as close as possible to the central source. This task is particularly critical at the longer wavelengths, where the PSF of the source emission dominates over a large fraction of the observed frames. 
Images of the continuum-free emission in individual lines have been obtained in the following way. First, a sub-cube covering a small wavelength range around the line of interest is created. Then we perform a linear fit through the continuum, estimated by carefully selecting a wavelength range on each side of the line free from any other emission feature. This continuum is then subtracted, pixel by pixel, in each frame of the sub-cube. In such a way we obtain, for each line,  a continuum-free sub-cube that is used on the one hand to construct total line emission maps, by integrating the emission in all the spectral elements covering the line profile, and, on the other hand, to build line velocity maps. For the latter, the central wavelength of each spectral frame has been converted to velocity by comparing it to the vacuum wavelength of the considered line, after performing the correction to the local standard of rest (LSR) velocities. The $V_{LSR}$ of HH46 IRS has been taken to be equal to the cloud velocity of $+5$ \kms\ following \cite{arce2013}. 

The linear fit of the continuum adjacent to the lines works relatively well for NIRSpec and for Channels 1 and 2 of MIRI. Above $\sim$ 10 \um\, the procedure leaves significant residual noise close to the source that compromises the identification of the line emission morphology. 
This is caused on the one hand by the residual fringes present at the pixel level due to the undersampling of the complex PSF morphology, and on the other hand by the significant increase of the continuum on-source level, that implies a relatively low line-to-continuum ratio for most of the lines.  In order to minimize this effect, we also tried a fit with a spline function, that better traces the continuum undulation, when present. This procedure improves the situation for some of the lines, allowing us to trace the emission a few pixels closer to the source. At the longest wavelengths, however, there is a limit where the noise on the counts in the continuum exceeds the line emission and therefore the line flux results to be under- or over-subtracted in adjacent pixels.

\section{The outflow} \label{outflow}

As shown in Fig. \ref{fig:FOV} the region covered by the JWST observations includes the base of the HH46 IRS atomic blue-shifted and red-shifted jet, and the wide angle cavity delineated by the CO ALMA observations. Here we separately discuss the results for these two components, that are sampled by different tracers, namely atomic forbidden lines for the jet and molecular emission for the cavity and the associated wide-angle wind.     

\subsection{The atomic jet}

\begin{figure*}[ht]
    \centering
    \includegraphics[width=1\textwidth,keepaspectratio]{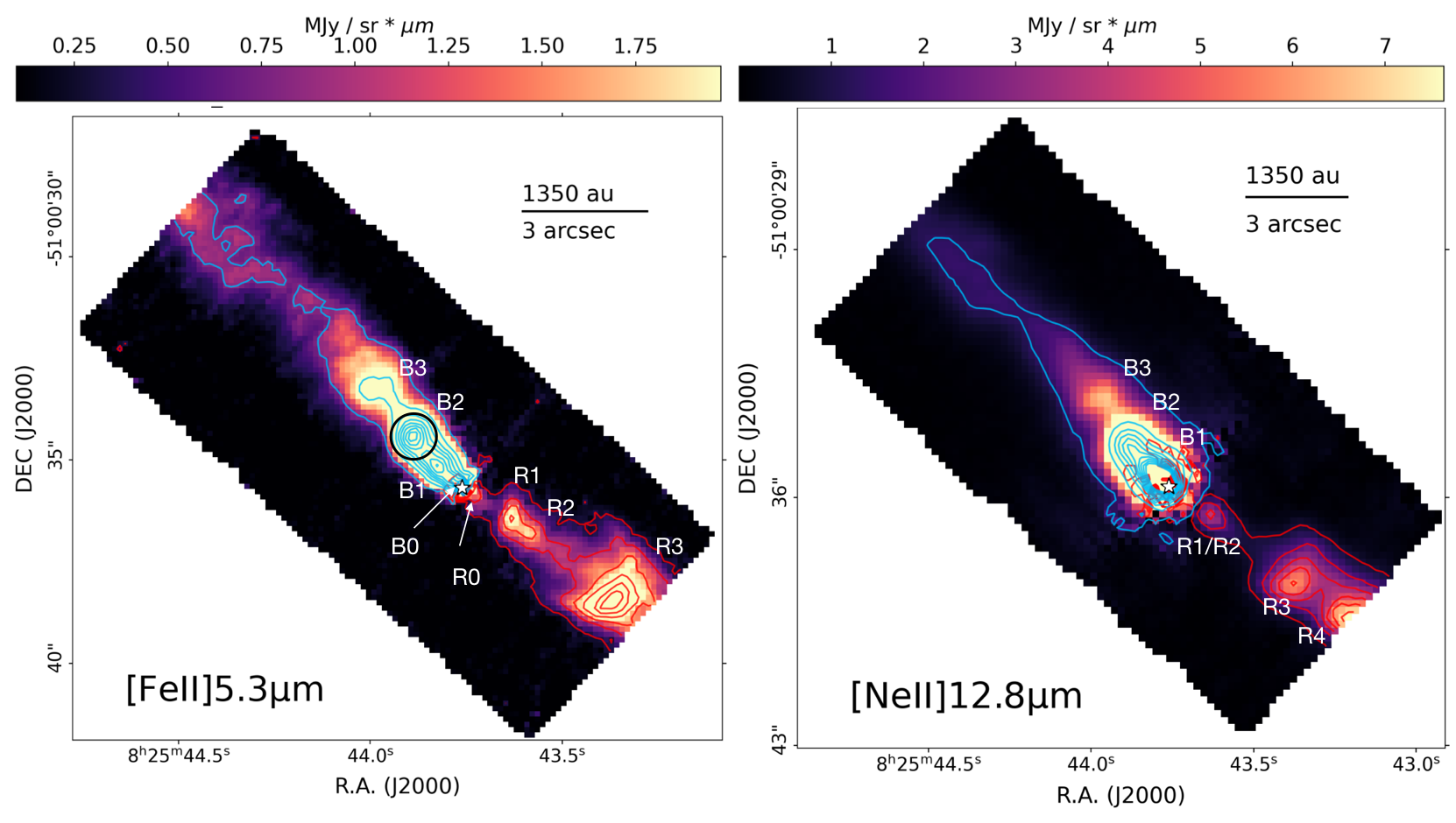}
    \caption{MIRI continuum-subtracted images of the emission in \feii\ 5.3\um\ (left) and \neii\ 12.8\um\ (right) lines. Blue and red contours delineate the blue-shifted and red-shifted outflow lobes, respectively. Contour levels are plotted in asinh scale and the values are as follow: \feii\ 
    blue  from 0.8 to 12.2 MJy\,sr$^{-1}$\,\um , \feii\ 
    red from 0.3 to 4.7 MJy\,sr$^{-1}$\,\um , \neii\ blue from 1.3 to 102.5 MJy\,sr$^{-1}$\,\um\ and \neii\ red from 0.75 to 18.7 MJy\,sr$^{-1}$\,\um . Bright emission knots are identified in the blue- (B0-B3) and red-shifted (R0-R3) jet. The black circle in the 5.3\um\ image indicates the circular aperture of 0\farcs 4 in radius from which the spectrum shown in Fig. \ref{fig:spettri_jet} has been extracted.
    The star indicates the position of the HH46~IRS source. 
    }
    \label{fig:MIRI_lines}
\end{figure*}

The MIRI spectral line images, covering a large field of view at wavelengths that suffer only limited extinction, allow us to get an understanding of the jet structure that is much improved with respect to previous optical/IR observations.  
Fig. \ref{fig:MIRI_lines} shows the continuum subtracted emission maps in the two brightest detected lines, namely \feii\ 5.3 \um, and \neii\ 12.8 \um , where the main emission peaks are indicated. The knots closest to the source (B0/R0) are detected at a distance of $\sim$ 0\farcs 25.

These two lines represent transitions to the ground state of the corresponding ions, excited up to levels with similar upper level excitation energies ($\sim$ 1000-2000 K). However, the ionisation potential (I.P) of the two species is significantly different, i.e. 7.9 eV for \feii\ and 21.56 eV for \ion{Ne}{2} (see Table \ref{table:atomic_lines}). Therefore, the detection of [\ion{Ne}{2}] lines indicate that the jet plasma is highly ionized.

\begin{figure}[ht]
    \centering
    \includegraphics[width=0.5\textwidth,keepaspectratio]{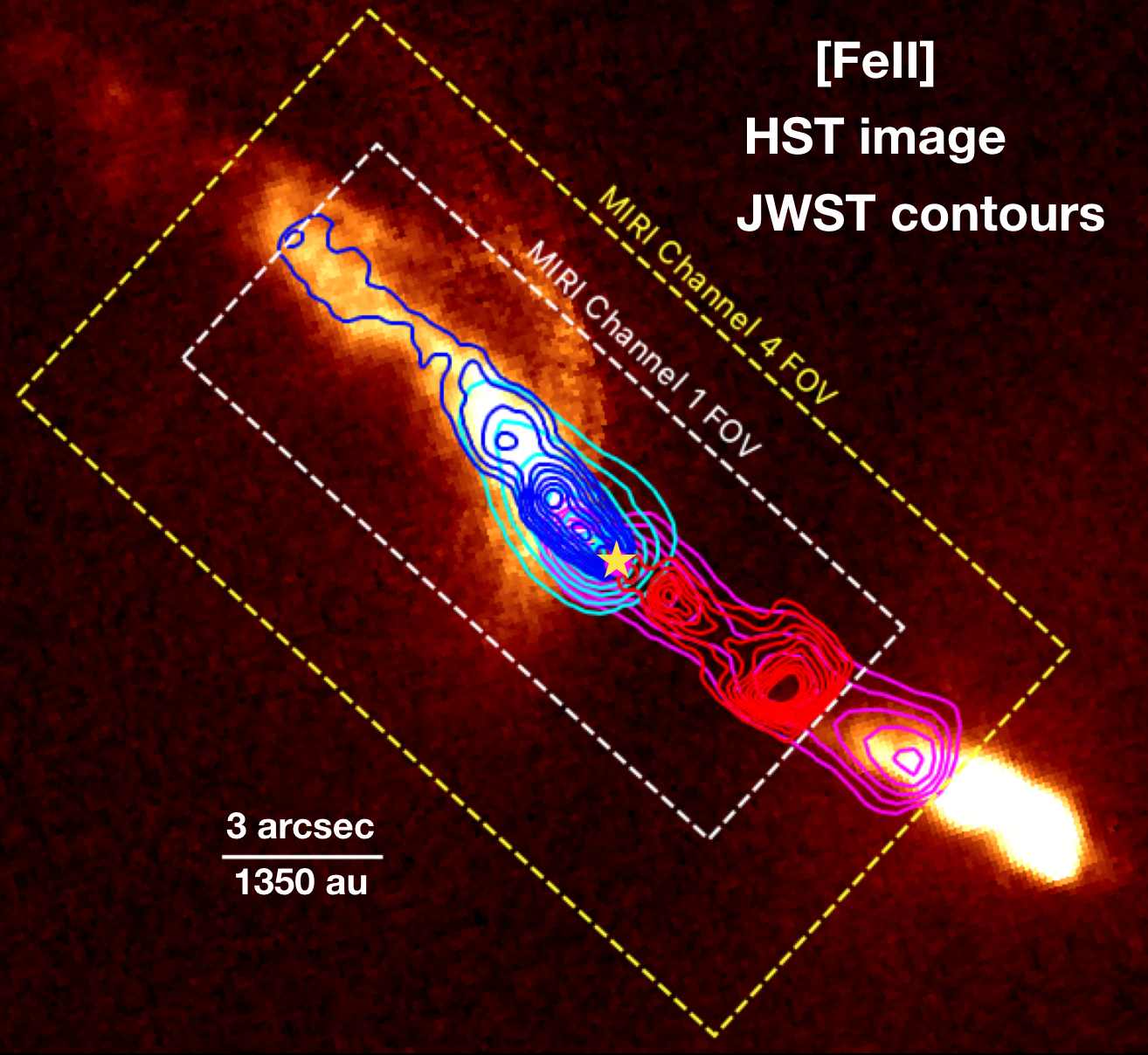}
    \includegraphics[width=0.4\textwidth,keepaspectratio]{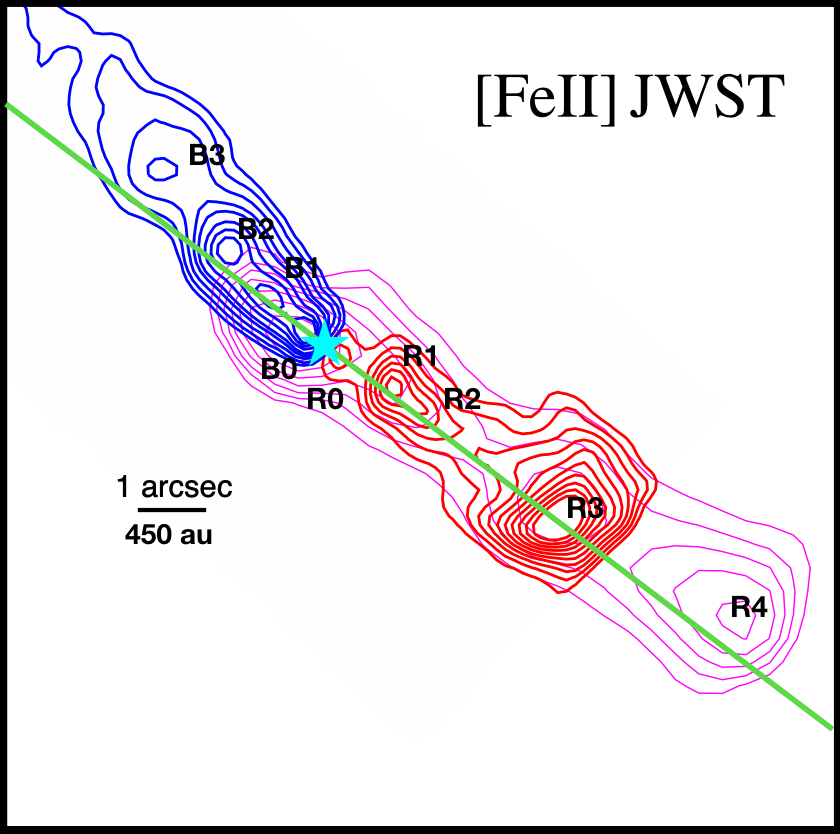}
    \caption{\textbf{Left:} Contours of continuum subtracted MIRI images of the \feii\ lines at 5.3 and 26 \um\ are overlaid on a continuum-subtracted HST image in \feii\  1.64 \um\ from \cite{erkal2021}. The MIRI \feii\ emissions are separately integrated into blue-shifted (blue 5.3 \um , cyan 26 \um )
    and red-shifted (red 5.3 \um , magenta 26 \um ) components. The area mapped in Channel 1 and Channel 4 is highlighted. \textbf{Right:} Contours of the inner jet region in the \feii 5.3 \um\ and 26 \um\ lines (colour code as in the left panel). The main emission peaks are indicated. The green line delineates the direction of the jet axis connecting the central source with the innermost B0 and R0 knots, highlighting that the outer knots deviate from this axis with a mirror symmetry between the blue- and red-shifted lobes.} 
    \label{fig:comp_HST}
\end{figure}

\begin{figure*}[ht]
    \centering
    \includegraphics[width=1\textwidth,keepaspectratio]{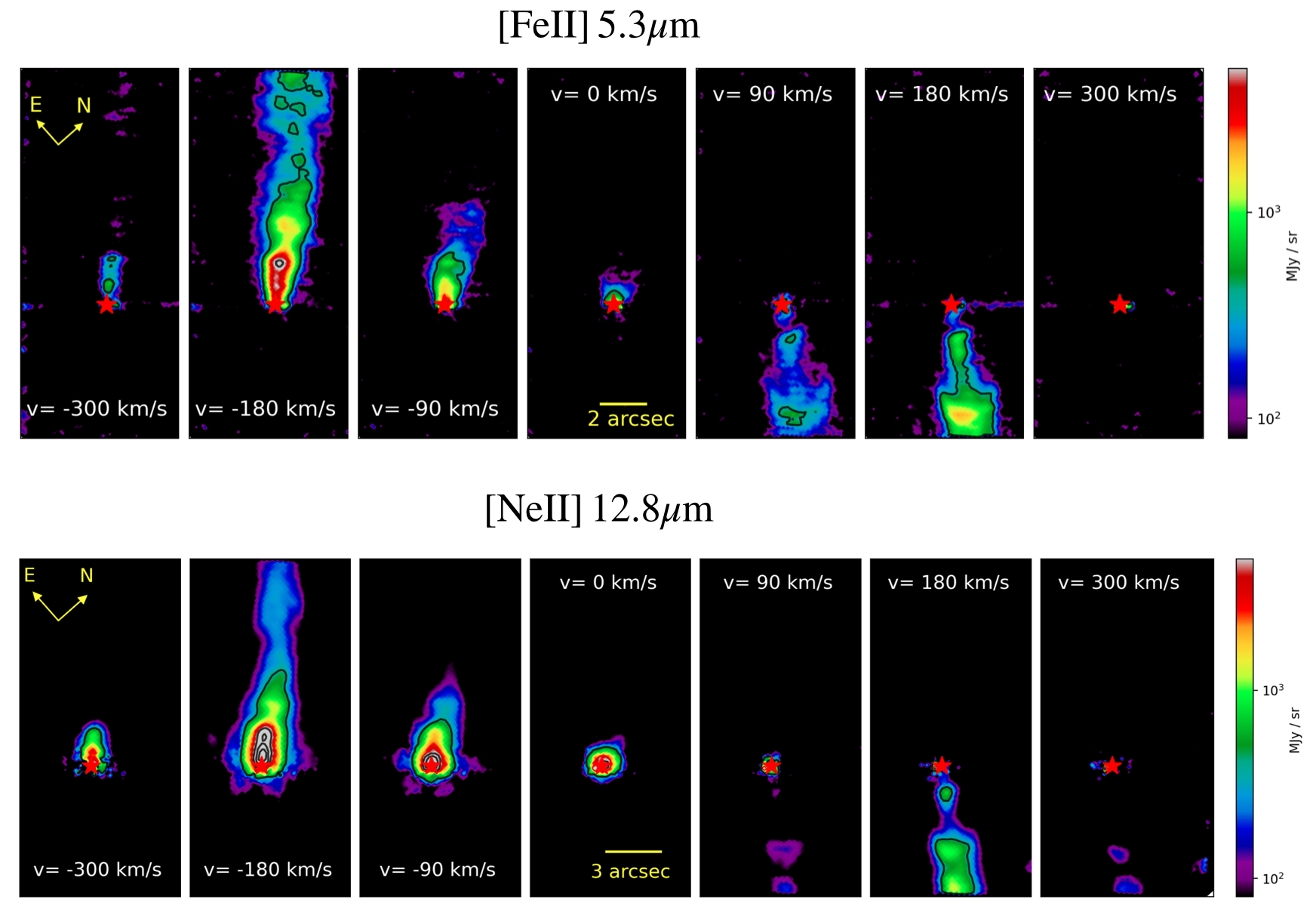}
    \caption{Velocity channel maps (integrated in 30 \kms\ bins)of the \feii\ 5.3 \um\ (top panel) and \neii\ 12.8 \um\ lines (bottom panel). The position of the central source is marked with a red star and all velocities are w.r.t. the systemic velocity. The images are rotated by -48.5 degrees NW direction. Contour levels are drawn from 400 to 5$\times$10$^4$ MJy\,sr$^{-1}$.}
    \label{fig:vel_map}
\end{figure*}

\begin{figure*}[ht]
    \centering
    \includegraphics[width=1\textwidth,keepaspectratio]{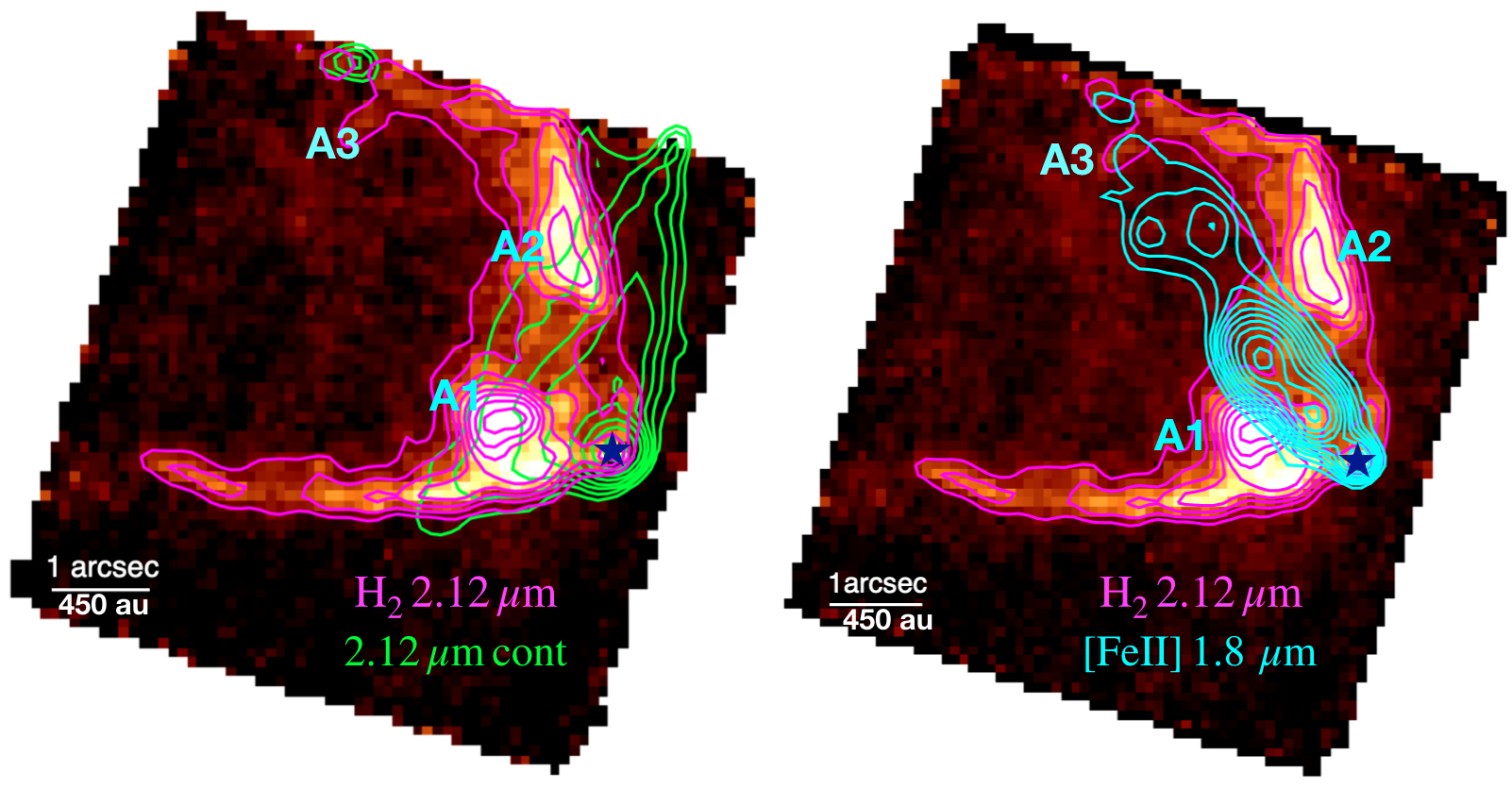}
    \caption{ Continuum-subtracted NIRSpec image of the integrated \htwo\, 2.12 \um\ emission of the blue-shifted inner region of the HH46IRS outflow. Overlaid contour levels goes from 0.025 to 0.3 MJy\,sr$^{-1}$\,\um\ in an asinh scale. 
    Contours of the 2.12 \um\, continuum emission (levels from 10 to 100 MJy\,sr$^{-1}$) are overlaid on the image in the left panel, while contours of the \feii\ 1.8 \um\ emission (levels from 0.025 to 0.3 MJy\,sr$^{-1}$\,\um), tracing the collimated jet, are overplotted on the right panel. The \htwo\ knots A1-A2 and the arc-shaped feature A3 located at the apex of the jet are labelled. The blue star indicates the position of the HH46~IRS source.}
    \label{fig:nirspec_image}
\end{figure*}

The most striking result appearing from these images is that at mid-IR wavelengths the red-shifted jet is clearly detected for the first time down to $\sim$ 90 au from the source. The 5.3 \um\ red-shifted emission shows a collimated jet broken in several emission peaks (R0-R2) driving an extended bow shock (R3) at a distance of about 4\asec. The \neii\ image, covering a larger field of view, detects an additional shocked knot (R4) located at a distance of about 7\farcs 3. 
To better highlight the comparison with previous observations, we show in Fig. \ref{fig:comp_HST} contours of the \feii\ 5.3 \um\ and 26 \um\ emission superimposed on a HST image of \feii\ 1.64 \um\ \citep{erkal2021}. The blue-shifted jet has a similar morphology in the \feii\ 1.64 \um\ and 5.3 \um\, lines, showing several emission peaks (named as B0-B3 in Fig. \ref{fig:MIRI_lines} ) plus a more diffuse extended emission at larger distances. In the red-shifted jet, the distinction into a jet plus a bow shock morphology is less evident for the 26 \um\ emission, due to the lower MIRI spatial resolution at these wavelengths. However, due to the larger FoV of MIRI Channel 4, the 26 \um\ emission covers larger distances, reconnecting to the external part of the jet detected at 1.64 \um .

Navarro et al. (2024, in preparation) estimate, from the analysis of the \htwo\ lines, that the visual extinction in the counter-jet is $>$ 35 mag in the inner regions and gradually diminishes to values $<$ 15 mag where the jet is again detected in the near-IR. 
These values are consistent with the estimated on-source A$_{\rm V}$ of roughly 45 mag by \cite{antoniucci2008}. 

The jet/counter-jet structure clearly does not follow a straight trajectory, and the detection of the counter-jet allows one to identify without ambiguity a mirror-like symmetry in the pattern followed by the flow, i.e. the jet trajectory in the red-shifted lobe mirrors the trajectory of the blue-shifted lobe. To better highlight this pattern, we show in the right panel of Fig. \ref{fig:comp_HST} the direction of the jet axis that connects the innermost B0/R0 knots: as the jet propagates, it progressively deviates from this direction and both the blue- and red-shifted knots are displaced towards the right of it.  
\cite{reipurth2000} explained the wiggling pattern of the HH46 blue-shifted jet as most likely due to orbital motion of the driving source around a companion. The mirror symmetry of the red-shifted jet supports this interpretation in contrast, for example, to precession of the jet axis \citep[e.g.][]{masciadri2002}. \cite{reipurth2000} identified a binary at a 0\farcs 26 separation in a 2\um\ HST image, but concluded that the separation was too wide to explain the HH46 wiggling observed at large distances. We also point out that the identified red-shifted knots R0-R3 are not displaced at exactly the same distance as the corresponding B0-B3 knots, with the exception of B0/R0. This could indicate that the location of shocked regions does not depend only on the time of the ejection events but also on the local environment encountered along the jet path. 
We also point out that the NIRSpec observations do not show clear evidence for the binary detected by \cite{reipurth2000}, the separation of which is instead well sampled by the NIRCam instrumental PSF (see Section 3.2.2 and relative discussion). 
We devote the analysis of the jet pattern in relation to the binary motion to a future dedicated paper.  

The rich atomic emission line spectrum of the HH46~IRS jet is presented in Figure \ref{fig:spettri_jet} of Appendix B, which shows the full MIRI and NIRSpec spectrum extracted from a 0\farcs4 radius aperture centered on the brightest blue-shifted jet knot (knot B2, see Fig. \ref{fig:MIRI_lines}).
In addition to the many \feii\ lines covering the entire wavelength range, we also detect emission of several ions of other abundant elements. 

Table \ref{table:atomic_lines} of Appendix C reports a list of the brightest detected atomic lines, including their excitation energies and the Ionisation Potential (IP) of the corresponding ions. Noticeably, the jet appears highly ionised, as we detected bright lines of species with I.P. up to 40 eV ([\ion{Ne}{3}]), and upper level energy up to 3e4 K, while lines of neutral atomic species, like e.g. [\ion{S}{1}] which dominates Spitzer spectra in other Class 0 jets \citep[e.g.][]{dionatos2010}, are here comparably weaker. We also detect, along the jet, several \hi\ recombination lines of the Pa, Br, Pf, and Hu series.

The [\ion{Ne}{3}] line emission, in particular, has a similar morphology as [\ion{Ne}{2}] shown in Fig. \ref{fig:MIRI_lines}, peaking close to the source (knot B0/B1) but extending up to more than 3000 au along the flow. The most likely origin for such an extended highly ionized gas is the action of high-velocity (v$> 80$ \kms\ ) shocks, able to produce ionizing UV/X-ray photons in the post-shock gas \citep[e.g.][]{hollenbach1989,hartigan1987}. Close to the source, a direct ionisation from high energetic photons from the central source could also be considered. For example, [\ion{Ne}{3}] has been detected at the base of a few CTT jets, through its UV line at 3869 \AA  \citep[e.g.][]{liu2016}. The observation of this line in the microjet of DG Tau has been explained by photoionisation due to hard X-rays produced in stellar flares, followed by slow recombination in the microjet. In HH46, the ratio [\ion{Ne}{2}]12.8 \um/[\ion{Ne}{3}]15.5 \um\ is about 20 on knot B1, which would be consistent with both scenarios of direct UV/X-ray photoionisation and shock excitation \citep{hollenbach2009} . However, the [\ion{Ne}{3}] recombination timescale is of the order of 1 yr \citep{glassgold2007}, while the flow timescale at the distance of the B1 knot or larger is $>$ 20 yrs \citep{hartigan2005}. Therefore a different source of ionisation, such as the UV/X-rays photons produced in energetic shocks, is needed to sustain the observed extended emission along the flow.

At the MIRI resolution, the kinematical structure of the jet is resolved as the jet velocity extends up to 300 \kms\ in either lobes \citep[see also][]{garcialopez2010}. 
Fig. \ref{fig:vel_map} presents velocity channel maps of the \feii\ 5.3 \um\ and \neii\ 12.8 \um\ emission. These have been obtained from our continuum subtracted sub-cubes for each line, and by interpolating the frames in order to have equally spaced interval of 30 \kms , which roughly correspond to sampling 1/3 of the MIRI spectral resolution in Channel 1.

From Fig. \ref{fig:vel_map} we see that close to the central source (i.e. within $\sim$ 3\asec) the jet displays a wide range of radial velocities, reaching up to 300 \kms . The jet terminal radial velocity at distances larger than $\sim$ 3\asec\ is within 180-200 \kms . The observed kinematical structure is roughly consistent with the velocity measured by \cite{garcialopez2010} for the \feii\ 1.64 \um\ line in long-slit spectra. Considering the jet inclination of $\sim$ 37\gradi\, with respect to the plane of the sky, the observed maximum radial velocity translates into a total jet velocity of about 380 \kms .

In spite of their different ionisation potentials, the range of velocities displayed by the \feii\ and \neii\ lines is similar. However, as can also be seen in Fig. \ref{fig:MIRI_lines}, \feii\ increases in brightness at larger distance from the source, while the \neii\  line gets fainter, indicating that the jet ionisation decreases with distance.
An interesting feature inferred from the \feii\ channel maps of Fig. \ref{fig:vel_map} is an apparent widening of the jet  
with decreasing velocity. 

To quantify this trend, we measured the jet width at about 1\asec\ from the source as the Full Width at Half Maximum (FWHM) of a Gaussian fit in the transversal direction, corrected for instrumental broadening ($\sqrt{FWHM_{fit}^2 - FWHM_{inst}^2}$) .
This width linearly decreases from $\sim$ 0\farcs 7 (315 au) in the -90\kms\ channel to 0\farcs 39 (175 au) in the -300\kms\ channel. 
This is in line with the trend found in atomic jets from CTT stars observed in the optical \citep[e.g.][]{maurri2014} and in molecular jets from Class 0 sources observed at millimeter wavelengths \citep[e.g.][]{podio2021}. The width of the jet is larger than inferred for Class II jets but similar to the width of Class 0 jets (~100-500 au at low velocity and ~50-250 au at high-velocity at ~450 au distance). The observed maximum width is in line with that measured by \cite{erkal2021} for the HST velocity integrated \feii\ 1.64\um\ image.
A detailed study of the excitation and physical properties of the jet, as a function of velocity, will be the subject of a specific paper in preparation.

\subsection{The molecular cavity and outflow}

\subsubsection{The small scale NIRSpec view}

Figure \ref{fig:nirspec_image} shows the NIRSpec image of the continuum-subtracted \htwo\ 2.12 \um\ emission with contours of the 2\ \um\ continuum emission (left panel) and of the \feii\ 1.81 \um\ emission (right panel) overlaid. 
The images clearly show that the atomic and molecular emission present very different spatial distributions, as also highlighted in \cite{birney2023}. In particular, we identify three main components for the \htwo\ extended emission delineating the edges of a parabolic cavity, two emission peaks, that we name A1 and A2, displaced at both sides of the atomic jet, and an arc-shaped structure (A3) located at the tip of the jet. Residual \htwo\, emission is observed also on-source, as well as diffuse emission inside the cavity. 
The comparison with the 2 \um\ continuum, which shows extended emission scattered in the reflection nebula, indicates that the \htwo\, emission cone lies inside the cavity, at least for the north side.  
Part of the \htwo\ emission along the cavity could be 
due to scattered line emission originating close to the source \citep[see e.g.][]{birney2023}. In addition to scattering, the \htwo\ emission along the cavity might originate from shocks caused by the impact of a wide angle flow on the cavity wall or from photo-evaporation \citep[e.g.][]{agra2014}. 

\begin{figure}[ht]
    \centering
    \includegraphics[width=0.6\textwidth,keepaspectratio]{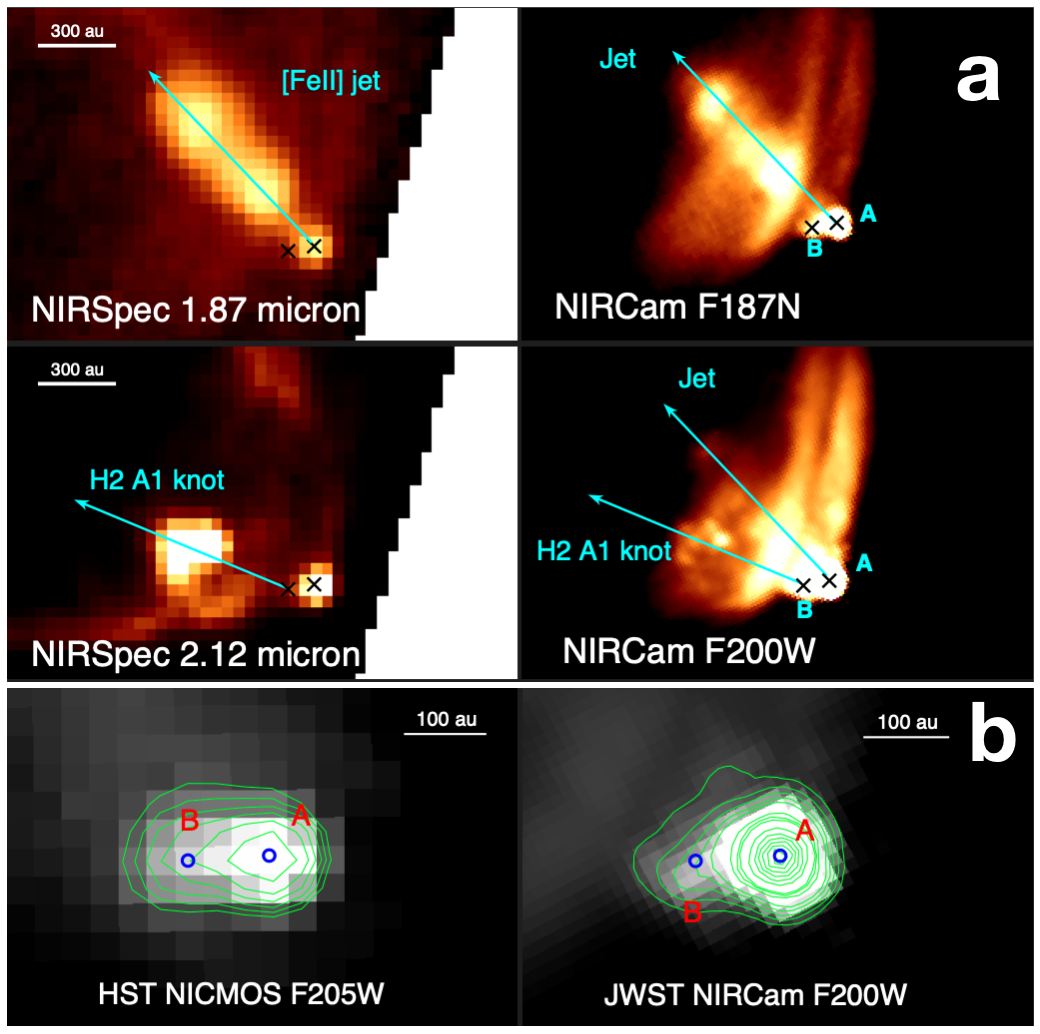}
    \caption{\textbf{(a)}: NIRSpec images of line+continuum emission at 
    1.87 and 2.12 \um\, (left) are compared with archival NIRCam images in the F187N and F200W filters (right). The NIRCam images (spatial resolution 31 mas/pixel) reveal a secondary source (B) located at about 0\farcs 23 from the 
    primary (A), i.e. about the position where \cite{reipurth2000} 
    reported the detection of a companion. The arrows indicate the direction of the jet axis driven by source A, and the \htwo\ A1 arc-shaped knot apparently pointing towards source B.
    \textbf{(b)}: Comparison of a HST image in 
    the F205W filter of the HH46 IRS binary taken in 1998 with the 
    NIRCam F200W image taken in 2023. Blue circles in both images indicate the position of the sources A and B in the HST image. Source B appears to have rotated around source A by about 13\gradi\ counterclockwise.
}
    \label{fig:binary}
\end{figure}

The location of the two \htwo\ peaks A1 and A2 might suggests at first sight that they are part of a wide angle molecular wind around the atomic jet, brighter at the edges due to limb-brightening. However, knot A2 is not oriented in the direction of the source and could be instead due to gas laterally compressed and entrained by the jet. 
The location of the A3 feature at the apex of the jet, and its arc shaped  morphology suggest that it could consist of material entrained as the jet is expanding. The shell morphology of knot A1, on the other hand, resembles that of an expanding bubble ejected directly from the source. The misalignment of this shell with respect to the jet axis is however puzzling
and difficult to explain by environmental effects only, leaving room to the hypothesis that this structure is driven by the secondary component of the binary system \citep{reipurth2000}, that we do not resolve with NIRSpec. 

\subsubsection{The binary at the origin of the two outflows ?}

To investigate this possibility, we used two images taken with NIRCam in the F187N and F200W filters (DDT program 4441), at a spatial resolution of 31mas/pixel. In Fig. \ref{fig:binary} we compare the NIRSpec spectral images at 1.87\um\ and 2.12\um\, with a section of the NIRCam images covering the same region. Noticeably the NIRCam images show a weaker emission peak at a distance of 0\farcs23 from the main source, and roughly along the same direction where \cite{reipurth2000} reported the position of the companion. 
The lower panel of Fig. \ref{fig:binary} directly shows the comparison between the NIRCam and HST images, the latter taken about 25 yrs earlier, in a similar filter. Noticeably the secondary companion, called source B by \cite{reipurth2000}, has changed its PA with respect to the primary A by about 13 degrees counter-clockwise, implying a binary period of the order of several hundreds years.

The NIRCam F187N narrow filter clearly detects the jet driven by the primary source A. It also shows that the emission from the reflection nebula is divided in two parts by a dark lane, maybe a region at a higher extinction, that also obscures the initial part of the jet. The NIRCam F200W filter covers several \htwo\ lines and also a few \feii\, lines. 
Due to the wide band and the presence of the strong nebulosity, the \htwo\ knots and the jet are only barely visible. In spite of that, 
knot A1 is well resolved and shows several sub-structures not resolved by NIRSpec, probably internal knots as sometimes are observed in more extended bow-shocks from YSO outflows. The image clearly shows that the jet emerges from source A with a PA of $\sim$ 43\gradi\, while the orientation of the A1 arc-shaped emission (PA $\sim$ 65\gradi ), suggests that it could instead be driven by source B. 
A rather similar situation has also been observed in the Class I binary TMC-1 by Tychoniec et al. (submitted). An analysis of the proper motion of the A1 structure, enabled by future JWST observations, will be needed to confirm this hypothesis.

\begin{figure*}[ht]
    \centering
    \includegraphics[width=1.0\textwidth,keepaspectratio]{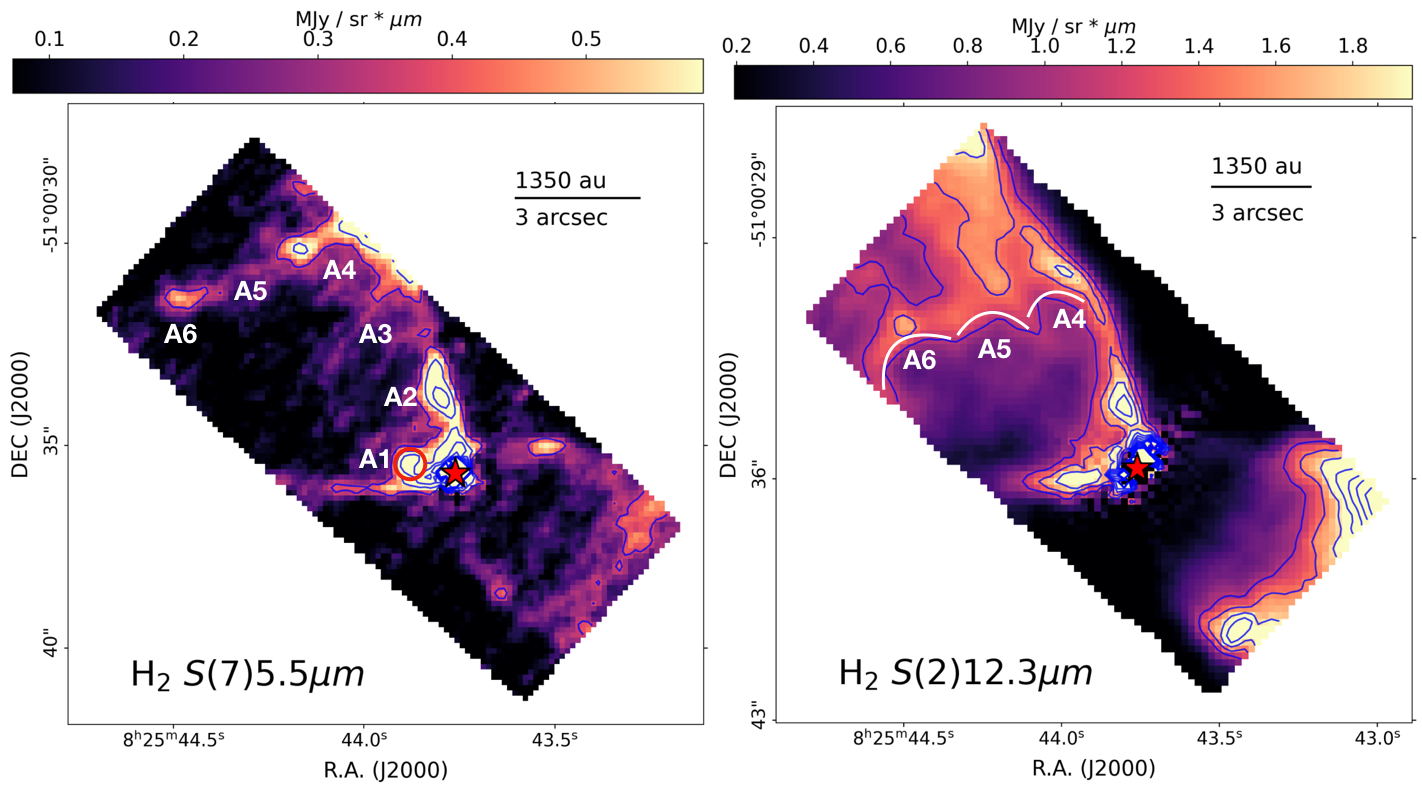}
    \caption{MIRI continuum-subtracted images of the emission in the \htwo\ 0--0 S(7) line at 5.5 \um\ (left) and 0--0 S(2) line at 12.3 \um\ (right). A Gaussian smoothing with $\sigma$ = 0.5 has been applied at both images. Contour levels are drawn in an asinh scale as follow: 5.5 \um\ from 0.36 to 2.42 MJy\,sr$^{-1}$\,\um , 12.3 \um\ from 1.07 to 6.65 MJy\,sr$^{-1}$\,\um . The main \htwo\ knots (A1-A2) and arc-shaped features (A3-A6) are indicated. The red circle indicates the aperture (0\farcs 4 in radius) from which the spectrum shown in Fig. \ref{fig:spettri_H2} has been extracted }
    \label{fig:MIRI_image}
\end{figure*}

\begin{figure}[ht]
    \centering
    \includegraphics[width=0.5\textwidth,keepaspectratio]{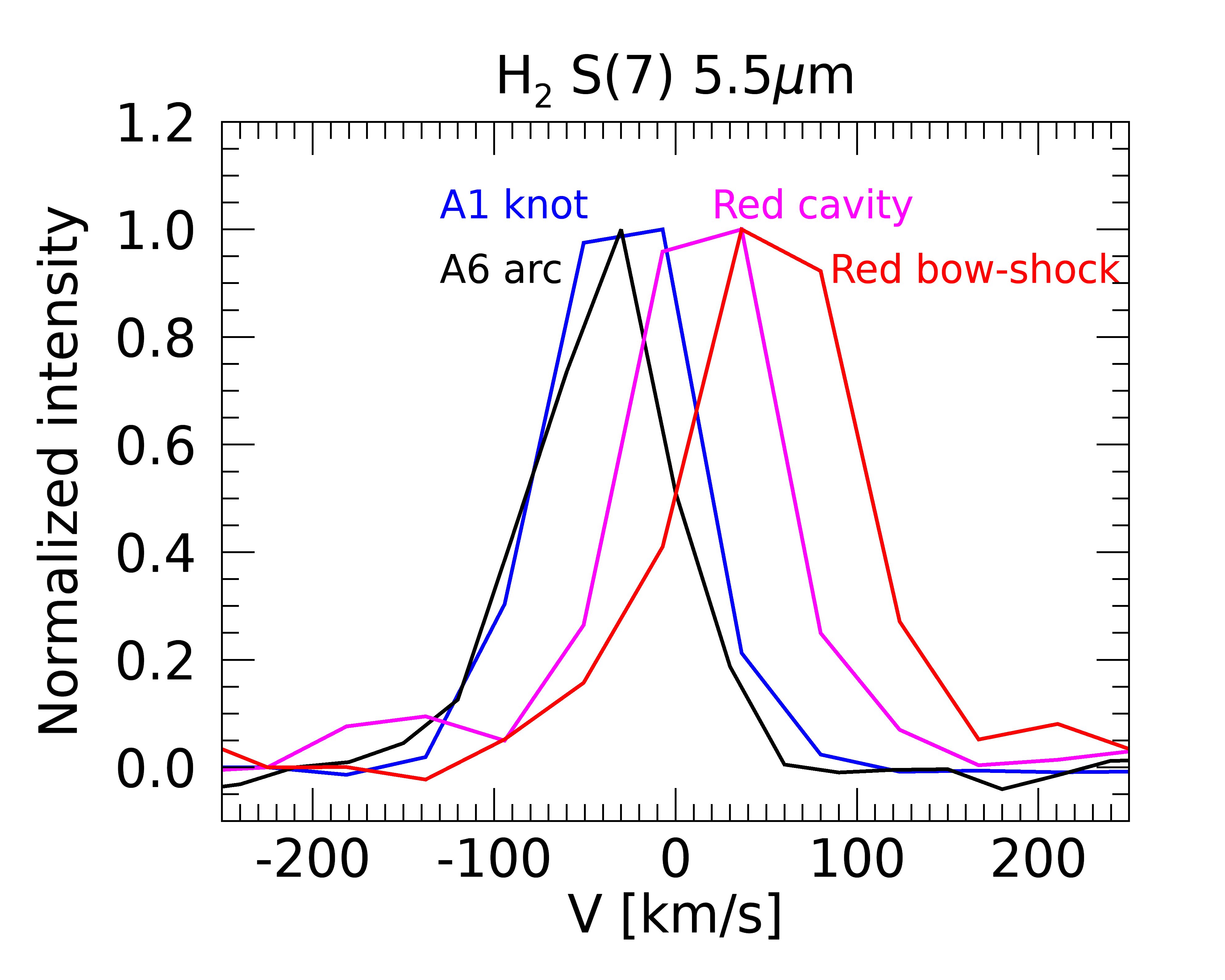}
    \caption{Spectra of the \htwo\ 5.5 \um\ line extracted in selected positions of the blue-shifted and red-shifted outflow. }
    \label{fig:h2_profiles}
\end{figure}

\begin{figure*}[ht]
    \centering
    \includegraphics[width=0.7\textwidth,keepaspectratio]{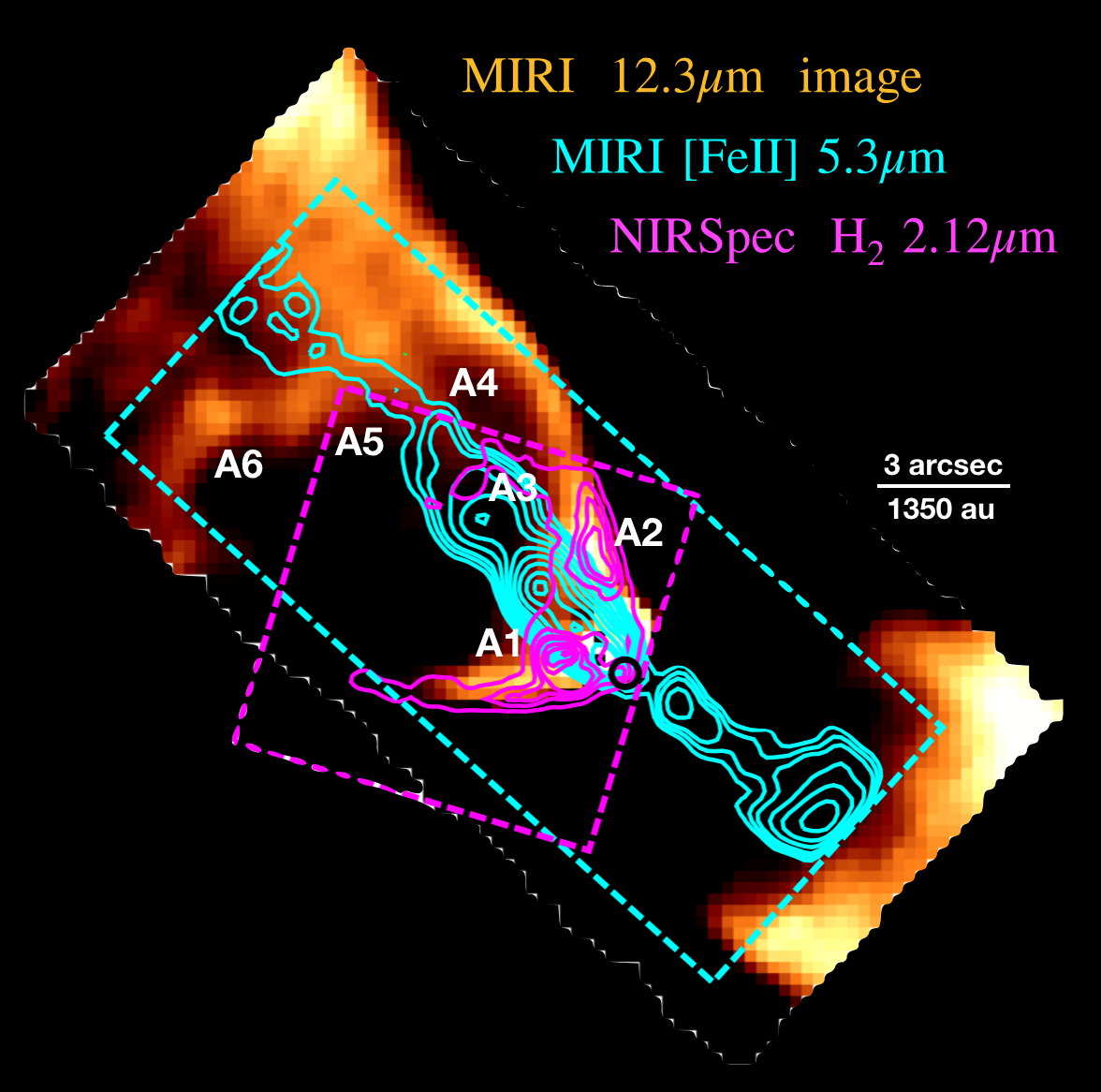}
    \caption{Continuum subtracted image of the \htwo\ 0--0 S(2) 12.3 \um\ line with overlaid contours of the \htwo\ 1--0 S(1) 2.12 \um\ observed by NIRSpec (magenta) and \feii\ 5.3 \um\ emission from the jet (cyan). Labels indicate the A1-A6 \htwo\ knots and arcs discussed in the text. The figure shows the similar morphology displayed by the rotational and ro-vibrational lines. It also suggests that the extended \htwo\ emission in the red-shifted outflow corresponds to a large bow-shock driven by the atomic jet. In contrast, the arc-shaped structures A5-A6 on the blue-shifted side are not oriented along the jet axis.} 
    \label{fig:global}
\end{figure*}

\begin{figure*}[ht]
    \centering 
\includegraphics[width=0.8\textwidth,keepaspectratio]{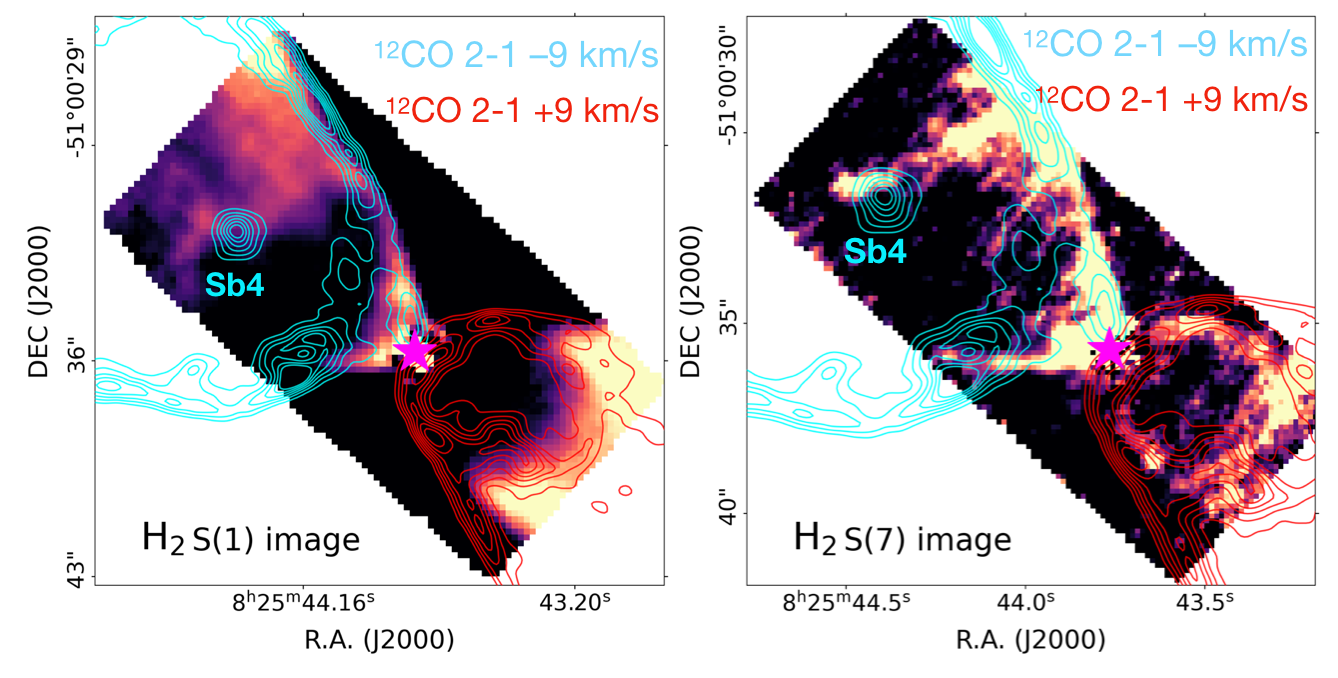}
    \caption{MIRI images of the emission in the \htwo\ 0--0 S(1) and S(7) lines with contours of the ALMA $^{12}$CO 2-1 emission in the velocity channels at V$_{LSR} \pm 9$ \kms (integrated within 1\kms). CO contour levels are drawn in a linear scales and have the following values: red-shifted from 0.05 to 0.24 K\,km\,s$^{-1}$, blue-shifted from 0.02 to 0.11 K\,km\,s$^{-1}$. The resolution of the ALMA observations is 0\farcs67$\times$0\farcs48.}
    \label{fig:H2_CO}
\end{figure*}

\subsubsection{The large scale MIRI view}

The MIRI observations allow us to study the distribution of the \htwo\ emission at larger scales, and to infer if molecular gas at low excitation, traced by the \htwo\, pure rotational lines covered by MIRI, displays a different morphology with respect to the warmer gas probed by the near-IR lines. 
Fig. \ref{fig:MIRI_image} shows the continuum-subtracted maps of the \htwo\, 0--0 S(7) line at 5.51 \um\ and 0--0 S(2) line at 12.28 \um\ (hereafter 5.5 and 12.3 \um). 
The morphology of the S(7) 5.5 \um\ emission is very similar to that of the 2.12 \um\ line for the region in common, and the features A1, A2, and A3 are here also well identified. In addition, on larger scales other arc-shaped features can be seen in the blue lobe delineating the edges of additional emission shells. 
The map of the S(2) line at 12.3 \um\, better shows three blue-shifted shells asymmetrically displaced with respect to the axis of the cavity, named A4, A5 and A6, together with bright diffuse emission that completely fills the cavity. 
In the red-shifted part of the outflow \htwo\ emission is observed along the cavity but at a lower SNR, mainly due to extinction. 
The red-lobe emission is more symmetric than the blue one, and here a very bright bow-shock, only barely covered by the S(7) map, largely dominates over the emission of the cavity in the S(2) emission. 

In spite of the low MIRI spectral resolution, some kinematical information can be retrieved for the observed components. Fig. \ref{fig:h2_profiles} shows the spectrum of the S(7) line extracted at different positions of the blue- and red-shifted lobes. Absolute wavelength calibrations do not allow us to give the radial velocity with an accuracy better than $\pm$ 30 \kms . The relative velocity shifts observed between different positions are however more reliable given the high SNR of the spectra. The blue-shifted emission at both the A1 peak and the A6 arc is consistent with a velocity -30 \kms.  
The red-shifted cavity appears shifted by about +15 \kms\ with respect to the blue outflow. In addition, we notice that the red-shifted bow-shock has a rather high radial velocity, reaching up to $\sim$ +60 \kms .  

Fig. \ref{fig:global} combines the \htwo\ S(2) emission with contours of the \htwo\ 2.12 \um\ and \feii\ 5.3 \um\ emission, for a global view of all the outflow components in the system. The morphology of the \htwo\ at 2.12 \um\ and 12.3 \um\, lines coincides along the cavity at the given resolution, thus we do not resolve any excitation stratification within the cavity. On the other hand, knot A1, which is bright in both the 2.12 \um\ and 5.5 \um\ lines, does not appear as prominent in the 12.3 \um\ image (as better highlighted in Fig. \ref{fig:MIRI_image}). For example, on knot A1 a 2.12 \um/12.3 \um\ intensity ratio equal to 3.7 is observed, while the same ratio is a factor of two lower on knots A2 and A3.
This indicates that knot A1 has a higher excitation with respect to the rest of the \htwo , as also confirmed by the analysis of Navarro et al. (2024, in preparation), which find on knot A1 an \htwo\ temperature up to $>$ 2000 K, in contrast to other \htwo\ emission knots in the region where the maximum temperature is typically in the range 1500-1900 K.

Figure \ref{fig:spettri_H2} of Appendix B shows the full MIRI and NIRSpec spectrum extracted from a 0\farcs4 radius aperture centered at the A1 knot (indicated in Fig. \ref{fig:MIRI_lines} and Fig. \ref{fig:MIRI_image}), while Table \ref{table:h2_lines} of Appendix C summarises the \htwo\ transitions detected. It can be seen that the spectrum is dominated by \htwo\ lines up to the v=3-2 ro-vibrational level, covering excitation energies from $\sim$ 400 K up to $\sim$ 15\ 000 K. A detailed analysis of the excitation conditions in the cavity and outflow will be presented in Navarro et al. (2024, in preparation). 

The comparison between the atomic jet and the molecular emission in Fig. \ref{fig:global} shows that the cavities and outflows in the blue- and red-shifted lobes appear very different. 
In the blue-shifted cavity, the \htwo\ emission arcs A4/A5 and A6 do not seem to be correlated to bow-shocks pushed by the jet, as they are not oriented in the jet direction (i.e. A6 arc), or apparently not associated with any jet knot. We point out that the jet is known to vary its orientation angle over time \citep{hartigan2005} due to its wiggling; however, the opening angle of the precession is about $\pm$ 15\gradi\, around the jet axis, thus much lower than, e.g., the angular distance between the jet axis and the A6 arc, which is around 25\gradi . In addition, the cooling time of the \htwo\ 12.3\um\ line is of the order of 70 yrs, in contrast with the more than 200 yrs ages of the jet knots located at further distance with proper motion deflected toward SE with respect to the present direction \citep[e.g. knots Jh5/Js7,][]{hartigan2005}.

It is therefore unlikely that the \htwo\ shells, as well as the misalignment of the jet with respect to the cavity, originate from the jet precession only. On the other hand, the A6 arc is roughly oriented in the direction of the A1 knot, (i.e. the IRS-B source, A1 and the apex of the A6 arc are aligned with an angle of $\sim$ 64\gradi ) and shares with it the same radial velocity, giving support to the possibility that the large scale \htwo\ arcs, (or at least the A6 arc), originate, as the A1 knot, from expanding shells ejected by the source B of the binary system. Precession and environmental effects might additionally contribute to shaping the overall morphology of the blue-shifted cavity as it appears in the \htwo\ images.
Noticeably there are no similar structures within the red-shifted cavity. Here, the large molecular bow-shock delineated by the \htwo\ S(2) 12.3\um\ emission is clearly correlated with the smaller size atomic bow-shock seen in \feii\ 5.3 \um . This latter shock may represent a Mach disk where the supersonic jet is decelerated and heated in a fast shock, which drives a slower non-dissociative bow shock, seen in \htwo , into the ambient molecular material. 

\subsubsection{Comparison between \htwo\ and CO ALMA observations}

The large asymmetries in structure and shape between the blue- and red-shifted cavities observed in \htwo\ emission, were also highlighted in the CO outflow observed with ALMA. \cite{arce2013} pointed out that the different morphology and opening angles of the blue- and red-shifted CO cavities could be due to different ambient densities encountered by the outflows in the two lobes. Given the location of HH46 at the edge of a globule, the blue-shifted flow travels more freely in a lower density environment, while the red-shifted outflow encounters a high-density region and thus entrains a larger amount of material. 

To better compare the morphology of the warm molecular emission with that of the cold CO gas, we overlay in Fig. \ref{fig:H2_CO} the ALMA $^{12}$CO 2-1 at velocity $\pm$ 9 \kms\ (cloud-corrected) with the \htwo\ S(7) 5.5 \um\ and S(1) 17 \um\ images. This latter image covers the largest area mapped with MIRI. As shown in \cite{zhang2019}, the $^{12}$CO 2-1 outflow presents several emission shells travelling at different velocities, with the higher outflow velocities extending farther from the source with respect to the lower velocities. We make here the comparison with a single CO low velocity channel as this roughly corresponds to the first observed structured shell, covering the region mapped with MIRI, and traces the cavity walls \citep[Figure 3 of][]{zhang2019}, for a direct comparison between the \htwo\ and CO cavities.
We see that the \htwo\ emission observed along the cavity is inside the CO emission in both lobes. This is particularly evident in the red-shifted cavity when comparing the \htwo\ S(7) and the CO morphology. This warm emission associated with the cavity could be driven by shocks or by photon-heating of the cavity walls. Heating of the outflow cavity walls by ultraviolet photons originating from the jet shocks and accretion disk was also proposed to explain the far-IR and sub-mm observations of high-J lines of CO, H$_2$O, and OH along the outflow \citep{vankempen2009,vankempen2010}.

The \htwo\ and CO emissions are instead spatially overlapping only in the northern part of the blue-shifted cavity. This could be due to the fact that in this direction the flow encounters a larger density and thus the gas cannot expand freely.
The \htwo\ shells in the blue cavity are not clearly associated with any CO emission shells, although a CO peak, called Sb4 in \cite{zhang2019}, is roughly coincident with the A6 arc. 
In the red-shifted lobe, the large \htwo\ bow shock is located further away with respect to the CO shell at 9 \kms . However, as previously pointed out, the CO emission presents shells at larger distance from the source as the velocity increases. Therefore the CO at 9 \kms\, is probably not kinematically linked with the \htwo\ bow shock, as this latter is travelling at much higher velocity. 

\cite{zhang2019} interpreted the CO shells morphology and velocity distribution as due to material swept up by an outbursting wide-angle wind. In this framework, the different shells could orginate from each intermittent mass ejection episode. \cite{zhang2016} in addition suggest that the red-shifted outflow can also be consistent with a bow-shock entrainment, where the jet  might expand and laterally push ambient material creating a wide cavity with low collimation. Indeed, models of pulsed jet-driven shells are qualitatively able to reproduce the morphological and kinematical features of the HH46 red-shifted cavity \citep{Rabenanahary2022}. The MIRI observations reinforce this interpretation as we observe for the first time the inner jet and the bow-shock, symmetrically located inside the cavity. 

As discussed before, for the blue-shifted outflow it is unlikely that jet-driven bow-shocks are the only driving mechanism at the origin of the CO entrainment and the complex system of \htwo\ shells seen with MIRI, although the spatial correspondence of a jet knot with the A3 shell seen in Figure \ref{fig:nirspec_image} suggests that the jet contributes to push material sideways in the NE direction. 
The evidence, however, that further out the outflow is not apparently associated with any jet knots, supports the hypothesis that the large cavity and associated \htwo\ emission originate from a 
wide angle molecular wind, that we suggest is launched by source B of the binary, although precession and interaction with the surrounding material might also play a role in shaping the overall cavity.

\section{The spectrum of the HH46~IRS protostar} \label{sec:protostar}

\begin{figure*}[ht]
    \centering
    \includegraphics[width=1\textwidth,keepaspectratio]{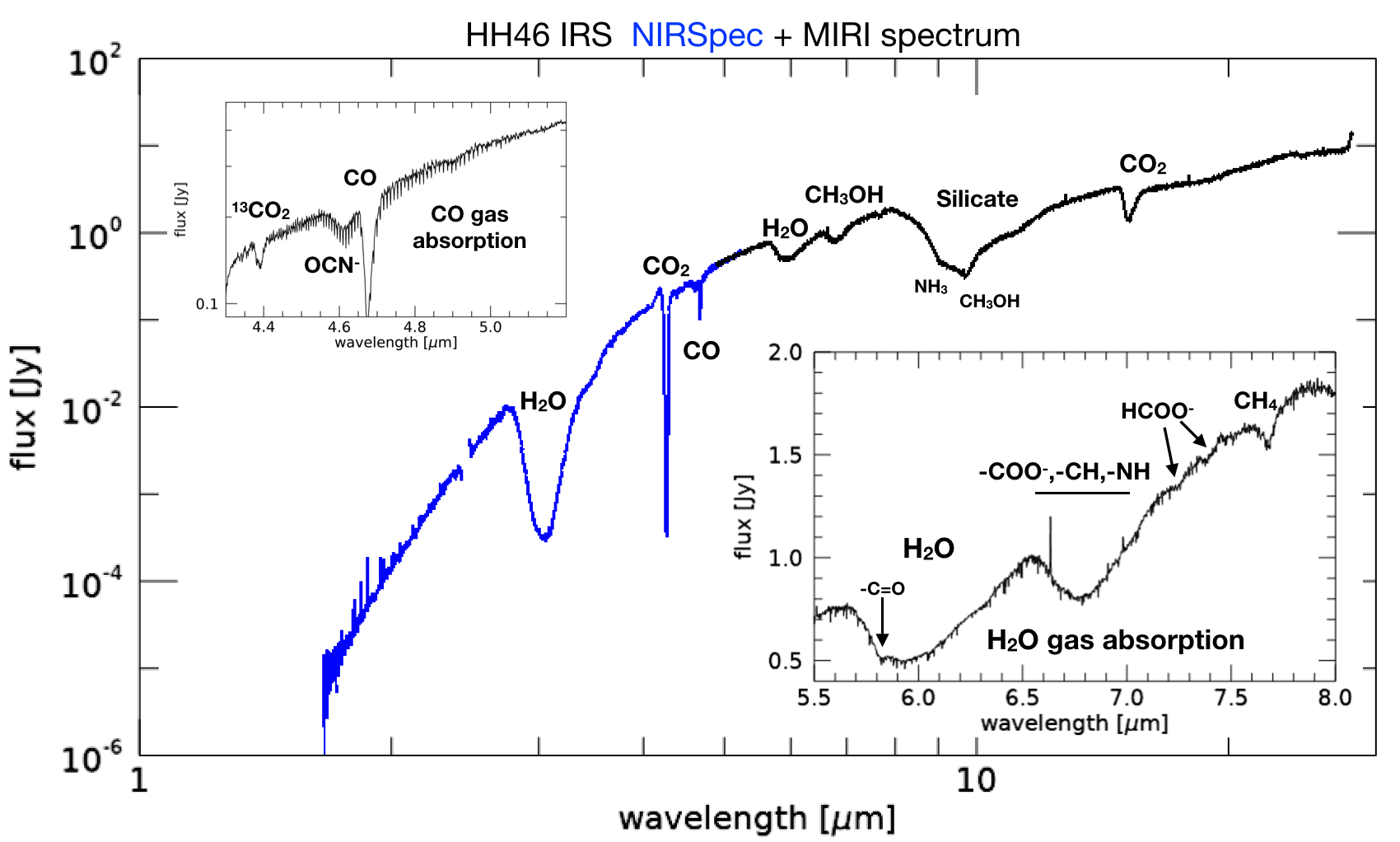}
    \caption{NIRSpec and MIRI MRS spectrum extracted at the position of HH46~IRS point source, with major solid-state features indicated. Insets show details of the 4.5 \um\ and 6.5 \um\ regions from the same spectrum where numerous rotational transitions of CO and H$_2$O are detected in absorption.}
    \label{fig:spettro_onsource}
\end{figure*}

Fig. \ref{fig:spettro_onsource} shows the complete NIRSpec + MIRI spectrum extracted at the source position as described in Sect. \ref{sec:extraction}. 
The spectrum steeply rises with wavelength and shows several absorption features due to ices and dust species. 
We clearly detect the 9.7 \um\, silicate band, as well as strong features of abundant molecules, such as H$_2$O, CO$_2$, CH$_3$OH, NH$_3$, OCN$^-$, some of them already observed in the Spitzer IRS spectrum of the source \citep{noriega2004,boogert2004}. In addition, thanks to the higher sensitivity and spectral resolution provided by JWST, we are able to detect weaker features due to less abundant and more complex organic molecules (COMs). As an example, the bottom inset of Fig. \ref{fig:spettro_onsource} shows the range 5.5-8 \um\ where, in addition to the broad 6 \um\ band attributed to the bending mode of H$_2$O, we also note a peak at 5.83 \um\ that might arise from the presence of molecules bearing a C=O bond, such as HCOOH/H$_2$CO \citep[e.g.][]{yang2022,rocha2023}. Also, the 6.78 \um\ band can be attributed to various compounds, including NH$_3$, NH$_4$$^+$ and other species bearing carboxyl and -CH groups. This feature also shows additional peaks at 7.256 \um\ and 7.376 \um\ that are attributed to HCOO$^-$, with possible contributions from more complex compounds, such as amides and aldehydes \citep[][]{urso2022}, that have been observed in other protostars \citep[e.g.][]{yang2022,mcclure2023,rocha2023}.

A peculiarity of this spectrum is the detection of many gas features seen in absorption instead of emission, at variance with the \jwst\ spectra of other low mass protostars \citep[e.g.][]{yang2022,federman2023}. 
 This can be seen for example in the upper inset of Fig. \ref{fig:spettro_onsource}, which highlight the spectral region around 4.5 \um , where absorption lines from the CO ro-vibrational transitions of the v=1-0 fundamental band are detected, and in the bottom inset covering the range 5.5-8 \um\, where many additional absorption lines, mostly from H$_2$O are present \citep[e.g.][]{gasman2023}. We remark that the CO v=1-0 lines are instead detected mostly in emission in the outflow, as can be seen in Fig. \ref{fig:spettri_jet} and Fig. \ref{fig:spettri_H2} of Appendix B. This indicates that the line absorption could arise in the dense molecular envelope surrounding the young stellar object, as found in a number of high-mass protostars by observations with the Infrared Space Observatory \citep[e.g.][]{helmich1996}, and recently also with \jwst\ (Francis et al. submitted). Alternatively, given the relatively high inclination of the putative disk (i.e. 53$^{\circ}$, if perpendicular to the jet orientation), the observed lines could also originate on the surface layer of an outer flared disk, that intercepts the mid-IR photons of the warm inner disk region. 
Strong absorption in the CO v=1-0 lines has indeed also been observed in some T Tauri stars and apparently correlates with the disk inclination \citep{banzatti2022}. 
A detailed analysis of the gas excitation conditions is required to disentangle these possibilities, which is deferred to a dedicated paper.

\begin{figure}[ht]
    \centering
    \includegraphics[width=0.5\textwidth,keepaspectratio]{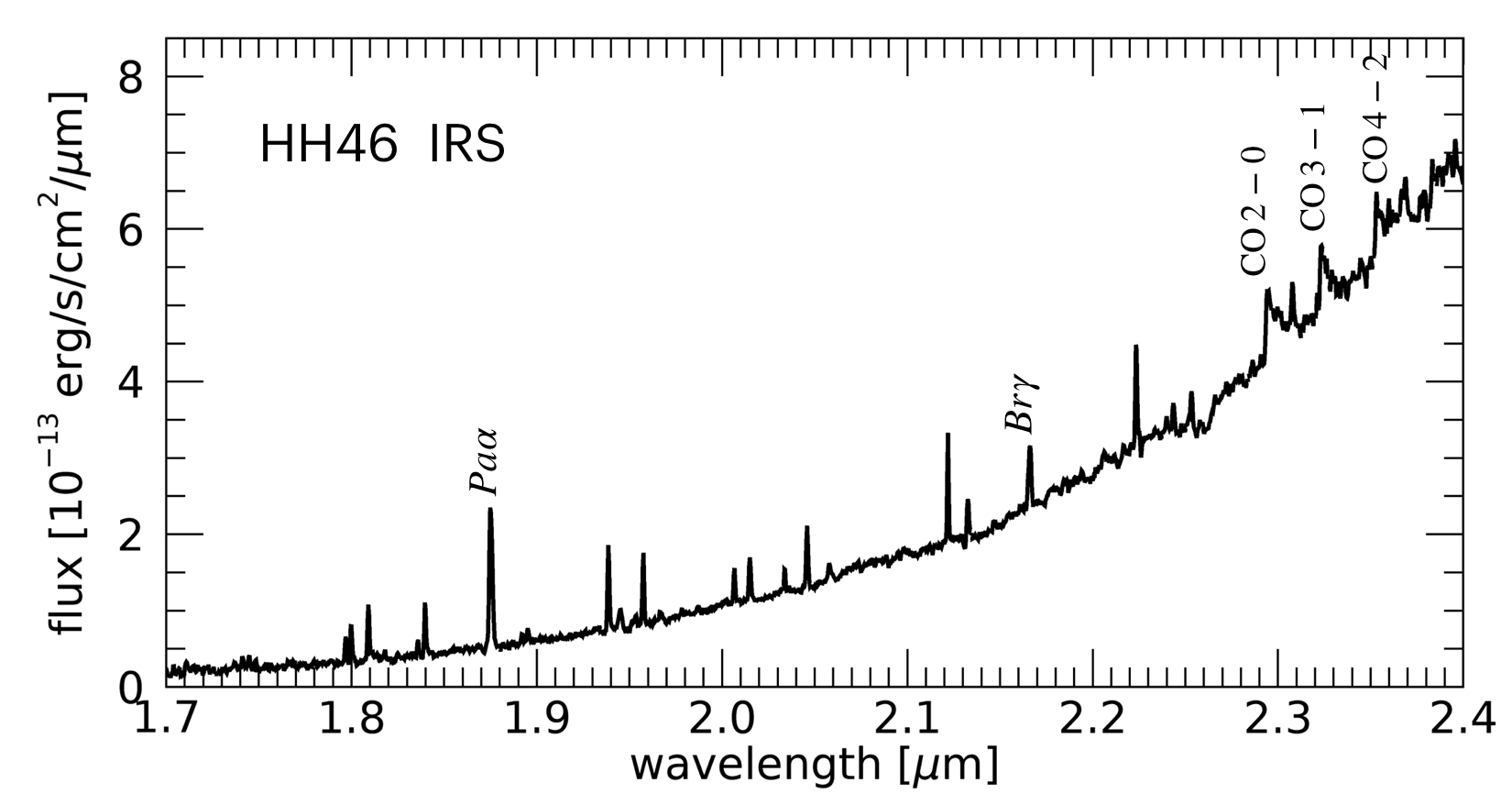}
    \caption{Section of the HH46~IRS NIRSpec spectrum around 2 \um . \hi\ lines and CO overtone emission lines are indicated.}
    \label{fig:star_NIR}
\end{figure}

\begin{figure*}[ht]
    \centering
    \includegraphics[width=1\textwidth,keepaspectratio]{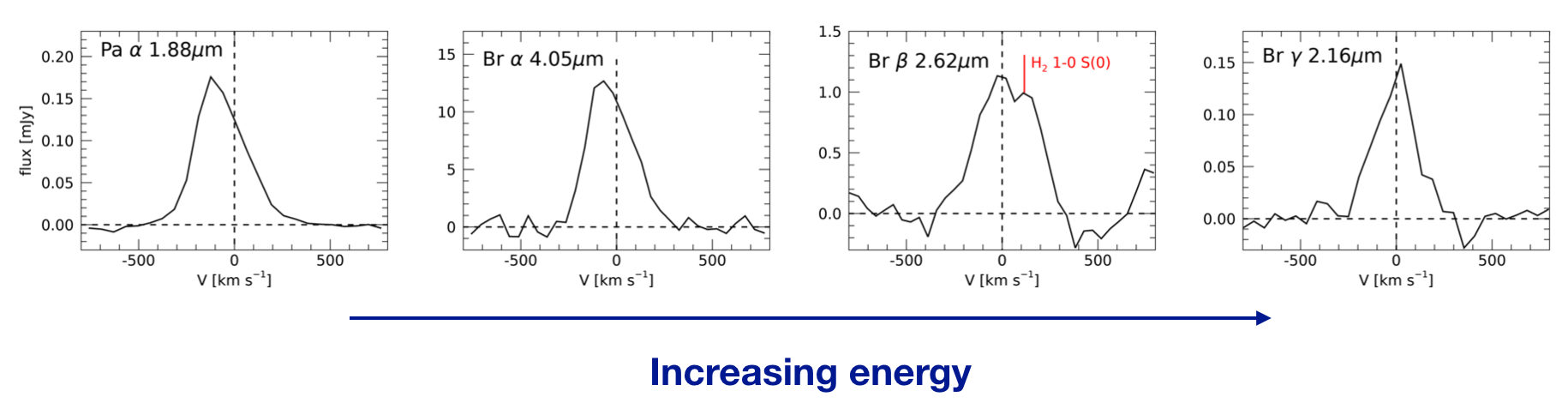}
    \caption{Hydrogen recombination lines detected in the HH46~IRS spectrum. The lines are ordered, from left to right, in increasing energy of the upper level. The emission peak shifts from being outflow dominated, at radial velocity $\sim -150$\kms , in the lines at lower energy, to being accretion dominated (V $\sim 0$ \kms) in the higher energy transitions. }
    \label{fig:HI_lines}
\end{figure*}

Fig. \ref{fig:star_NIR} shows a zoom of the NIRSpec spectrum around the 2\,\um\ range. This can be compared with the spectrum obtained from ground using 8-m telescopes \citep{antoniucci2008,birney2023}. Even with the larger SNR attained with NIRSpec with respect to the ground-based observations, no photospheric absorption lines \citep[e.g.][]{nisini2005,greene2018} are detected here, which implies that the IR line and continuum emission from the disk and accretion spots dominate and completely veil the stellar photosphere. 
The CO overtone bandheads ($\Delta v = 2$) are detected in emission, at variance with the lines of the fundamental vibrational band detected in absorption. These overtone emission lines are rarely detected from T Tauri disks, but are often displayed by highly accreting Class I sources. The detection of the CO bandheads implies kinetic temperature of a few thousands K and their emission likely originates from the hot gas in the innermost disk region. 

In the HH46~IRS spectrum we detected four \hi\
lines, namely Pa$\alpha$ at 1.88 \um , Br$\gamma$ at 2.16 \um , Br$\beta$ at 2.62 \um, and Br$\alpha$ at 4.05 \um . 
Pfund and Humphrey lines covered in the MIRI range and seen in the outflow (see Table \ref{table:atomic_lines}) are not detected on source at more than the 2$\sigma$ level due to the low line/continuum ratio. Fig. \ref{fig:HI_lines} shows the profiles of the detected lines ordered by increasing energy of the upper level. It can be noticed that the peak of the line shifts from about $-150$ \kms\ for the less energetic lines (Pa$\alpha$ and Br$\alpha$), to about 0 \kms\ for the lines at higher energy (Br$\gamma$ and Br$\beta$). This shows that the line emission shifts from being outflow-dominated to being accretion-dominated, depending on the excitation condition. The ionized jet at its base 
significantly contributes to the \hi\ emission of the low-level lines of each series, while higher-level lines of the same series are dominated by emission from the higher density gas of the accretion columns, that causes an increase of the line optical depth and thus a deviation from the optically thin Case B \citep[][]{antoniucci2017}. We point out, however, that the jet still contributes, even if at a lower level, to the emission of high excited lines such as the Br$\gamma$ line, as this emission at high velocity is spatially extended, as also shown in \cite{birney2023}. 

In low mass T Tauri stars, optical and IR hydrogen recombination lines are commonly used to derive the source accretion luminosity ($L_{acc}$), adopting empirical
relationships that correlate the \hi\ line luminosity with $L_{acc}$ \citep[e.g.]
[]{alcala2017}. Similar relationships have been tentatively found for lines in the mid-IR \citep[e.g.][]{rigliaco2015,salyk2013} which in principle would allow one to estimate $L_{acc}$ also for very embedded young stellar objects like the Class I sources. The results we find on HH46 show that some care needs to be applied when using different lines as independent proxy of the accretion luminosity in young sources with energetic jets, as the relative contribution of accretion and jet emission significantly changes from line to line. A detailed analysis about the accretion properties of HH46 IRS will be addressed in a future paper. 

\section{Conclusions}

This paper presents an overview of the results obtained with JWST MIRI-MRS and NIRSpec IFU observations of the Class I protostar HH46~IRS and its outflow, as part of the PROJECT-J Cycle 1 program 1706. 
We mapped a region of $\sim$ 6\farcs 6 in length with NIRSpec, and up to $\sim$ 20$^{\prime\prime}$ with MIRI, that includes the central protostar, its collimated jet and the associated molecular outflow and cavity. These IFU observations provide spectral maps covering the entire wavelength range from 1.66 to 28 \um\ at a spatial sampling between 0\farcs 1 to 0\farcs 3, and with a spectral resolution between 1500 and 3500. The highlights shown by these observations can be summarised as follow:

\begin{itemize}
\item  We detect for the first time both lobes of the atomic collimated jet to within $\sim$ 90 au from the source, inferring a visual extinction larger than 35 mag at the base of the red-shifted counter-jet. The jet displays plenty of forbidden lines of abundant ions, the brightest two being \feii\ at 5.3 \um\ and \neii\ at 12.8\,\um .

The jet is highly ionized, as testified by the detection of several \hi\ lines and transitions of ions such as the [\ion{Ne}{3}] 15.5\,\um\  and [\ion{Ar}{3}] 8.99\,\um. Such evidence indicates excitation by high velocity ($>$ 80 \kms) shocks, although for the emission closest to the source 
direct ionisation from UV/X-ray stellar photons cannot be excluded. No collimated molecular emission is detected along the jet.

\item We resolve the velocity structure of the jet, that shows radial velocities up to $\pm$ 300 \kms . Considering an inclination angle of 37\gradi\ , these translate into a total jet terminal velocity of $\sim$ 380 \kms . Similar terminal velocities are observed in lines at different excitation. We also observe a decrease of the jet opening angle with increasing velocity. The jet width at about 1\asec\ from the source linearly decreases from $\sim$ 315 au in the -90 \kms\ channel to 175 au in the -300 \kms\ channel. 

\item The trajectory of the jet and counter-jet display a mirror symmetry, indicative of orbital motion of the jet-driving source in a binary system. The morphology of the two lobes is however highly asymmetric, probably due to the different ambient medium in which the jet and counter-jet are travelling. 

\item Archival NIRCam images of the central region resolve the 0\farcs 23 binary system reported in previous HST NIR observations. The projected motion of the binary is seen here for the first time comparing the NIRCam and HST images, acquired in a time span difference of 25 yrs, and suggest a binary period of the order of several hundreds of years.

\item Spectral images of \htwo\ lines at different excitation outline a complex molecular flow, where the blue-shifted lobe shows bright emission along the cavity, \htwo\ peaks uncorrelated with the collimated jet, and several large-scale molecular arcs. The NIRCam observations support the hypothesis that the observed blue-shifted \htwo\ emission, with a morphology of an expanding shell, is driven by the companion. This evidence, together with jet precession and density gradients within the cavity could explain the large asymmetries observed between the jet, driven by the primary source, and the wide-angle molecular structures observed at larger distance.
The red-shifted lobe displays weaker \htwo\ emission associated with a cavity and a bright jet-driven bow-shock. The large asymmetry between the outflows in the two lobes are primarily due to the difference in the ambient conditions encountered by the outflow as it travels away from the source. 

\item The comparison between the MIRI images and a CO ALMA low-velocity map shows emission from warm \htwo\ gas that follows the CO cavity but that lies inside it. This warm emission from the cavity could be driven by shocks or by photon-heating of the cavity walls.
The flanks of the \htwo\ red-shifted bow-shock connect with the cavity edges, giving support to the scenario where the cavity is created by entrainment of jet-driven bow-shocks. This is at variance with the blue-shifted lobe, where the expansion of both the collimated jet and the wide angle molecular shells driven by the two sources concur to create the asymmetric CO cavity observed with ALMA. 
\item The spectrum of the non-resolved binary system shows deep structured ice bands and CO v=1-0 and H$_2$O v=1-0 gaseous lines in absorption, likely originating in the envelope or disk intercepting the warm mid-IR continuum. Several hydrogen recombination lines of the Paschen and Brackett series are observed on-source. Their emission peaks shift from being outflow-dominated to being accretion-dominated as the energy of the line increases.   
\end{itemize}

This paper shows the richness of information that can be gathered with JWST observations on a complex system such as HH46. We have in particular unveiled the central engine of the HH46~IRS outflow, showing that jet-entrainment and wide-angle winds from the two sources are both at play in shaping the large scale structure of the outflow. The obtained data demonstrate the power of JWST in the investigation of embedded regions around young Class I protostars, which remain elusive even at near-IR wavelengths. A series of forthcoming papers currently in preparation will examine in depth a number of topics highlighted in this article, including the excitation and dynamics  
of the \htwo\ outflow and of the atomic jet, as well as the nature of the gas and dust features of the source. 

\begin{acknowledgments}

This work is based on observations made with the NASA/ESA/CSA James Webb Space Telescope. The data were obtained from the Mikulski Archive for Space Telescopes at the Space Telescope Science Institute, which is operated by the Association of Universities for Research in Astronomy, Inc., under NASA contract NAS 5$-$03127 for JWST. These observations are associated with program \#1706 and can be accessed via \dataset[DOI: 10.17909/eav1-0619]{https://doi.org/10.17909/eav1-0619} 

PH, TG and HA acknowledge funding support from JWST GO program \#1706 provided by NASA through a grant from the Space Telescope Science Institute. We gratefully acknowledges the help of the Space Telescope Science Institute JWST Helpdesk, and in particular Jane Morrison, for her valuable suggestions on the NIRSpec data reduction. 
INAF co-authors acknowledge support from the Large Grant INAF 2022 “YSOs Outflows, Disks and Accretion: towards a global framework for the evolution of planet forming systems (YODA)” and from PRIN-MUR 2022 20228JPA3A “The path to star and planet formation in the JWST era (PATH)”.  
LP acknowledges the PRIN MUR 2022 FOSSILS (Chemical origins: linking the fossil composition of the Solar System with the chemistry of protoplanetary disks, Prot. 2022JC2Y93). EvD acknowledges the funding from the European Research Council (ERC) under the European Union’s Horizon 2020 research and innovation programme (grant agreement No. 291141 MOLDISK). SC gratefully acknowledges support from the Programme National de Physique et Chimie du Milieu Interstellaire (PCMI), cofunded by CNRS-INSU, CNES, and CEA, and from Observatoire de Paris (Action Fédératrice Incitative Univers Froid).  TPR acknowledges support from the ERC (grant agreement No. 743029 EASY). CFM is funded by the European Union (ERC, WANDA, 101039452). Views and opinions expressed are however those of the author(s) only and do not necessarily reflect those of the European Union or the European Research Council Executive Agency. Neither the European Union nor the granting authority can be held responsible for them. This paper makes use of the following ALMA data: ADS/JAO.ALMA \#2012.1.00382.S. ALMA is a partnership of ESO (representing its member states), NSF (USA), and NINS (Japan), together with NRC (Canada), NSC and ASIAA (Taiwan), and KASI (Republic of Korea), in cooperation with the Republic of Chile. The Joint ALMA Observatory is operated by ESO, AUI/NRAO, and NAOJ. This work benefited from the Core2disk-III residential program of Institut Pascal at Universite\'e Paris-Saclay, with the support of the program “Investissements d’avenir” ANR-11-IDEX-0003-01. 

\end{acknowledgments}

\vspace{5mm}
\facilities{JWST}
\facilities{HST}
\facilities{ALMA}




\newpage

\appendix

\section{The HH46~IRS extracted spectrum and comparison with Spitzer.}

\begin{figure*}[ht]
    \centering

    \includegraphics[scale=0.7]{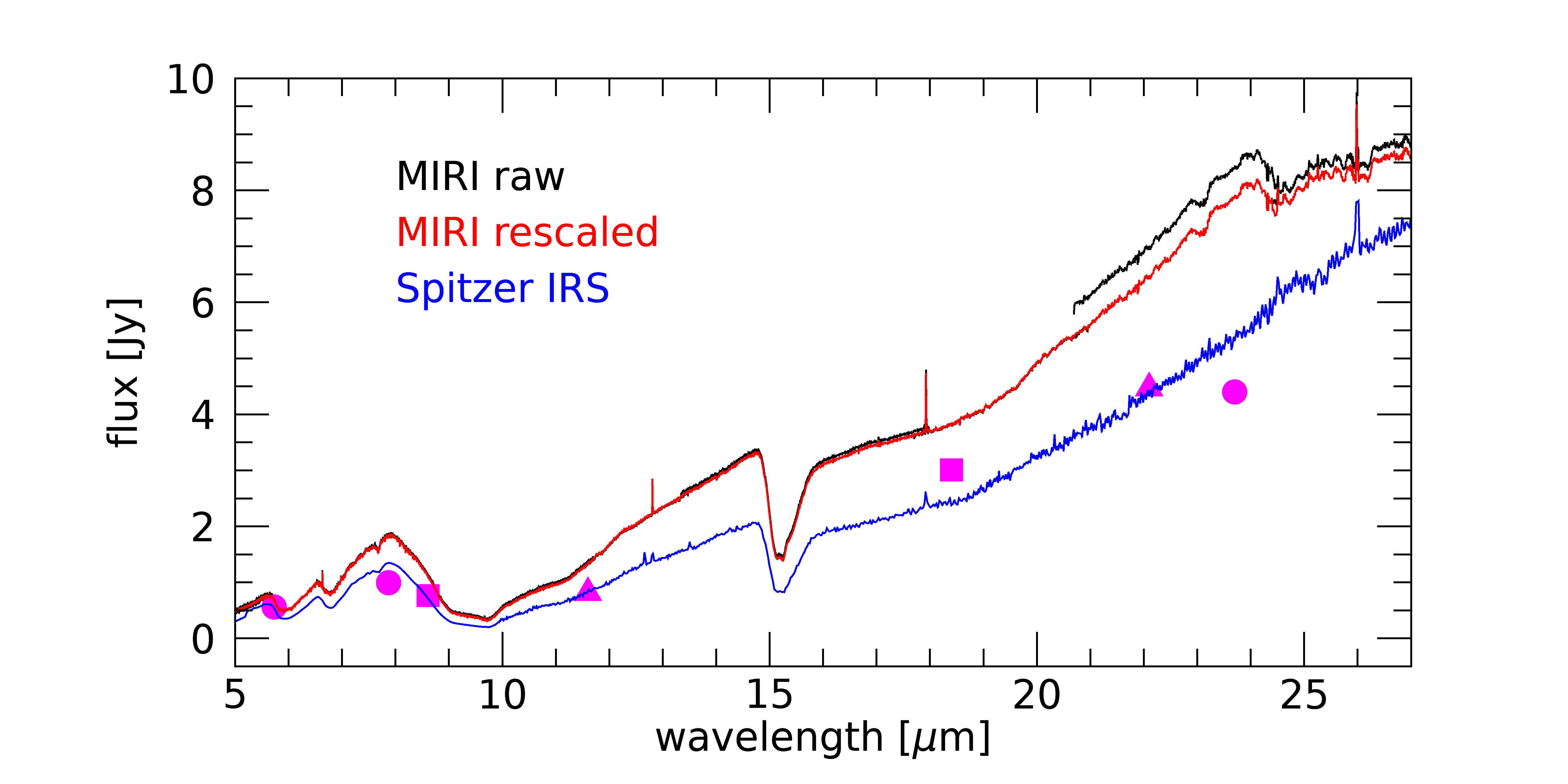}
    \includegraphics[scale=0.7]{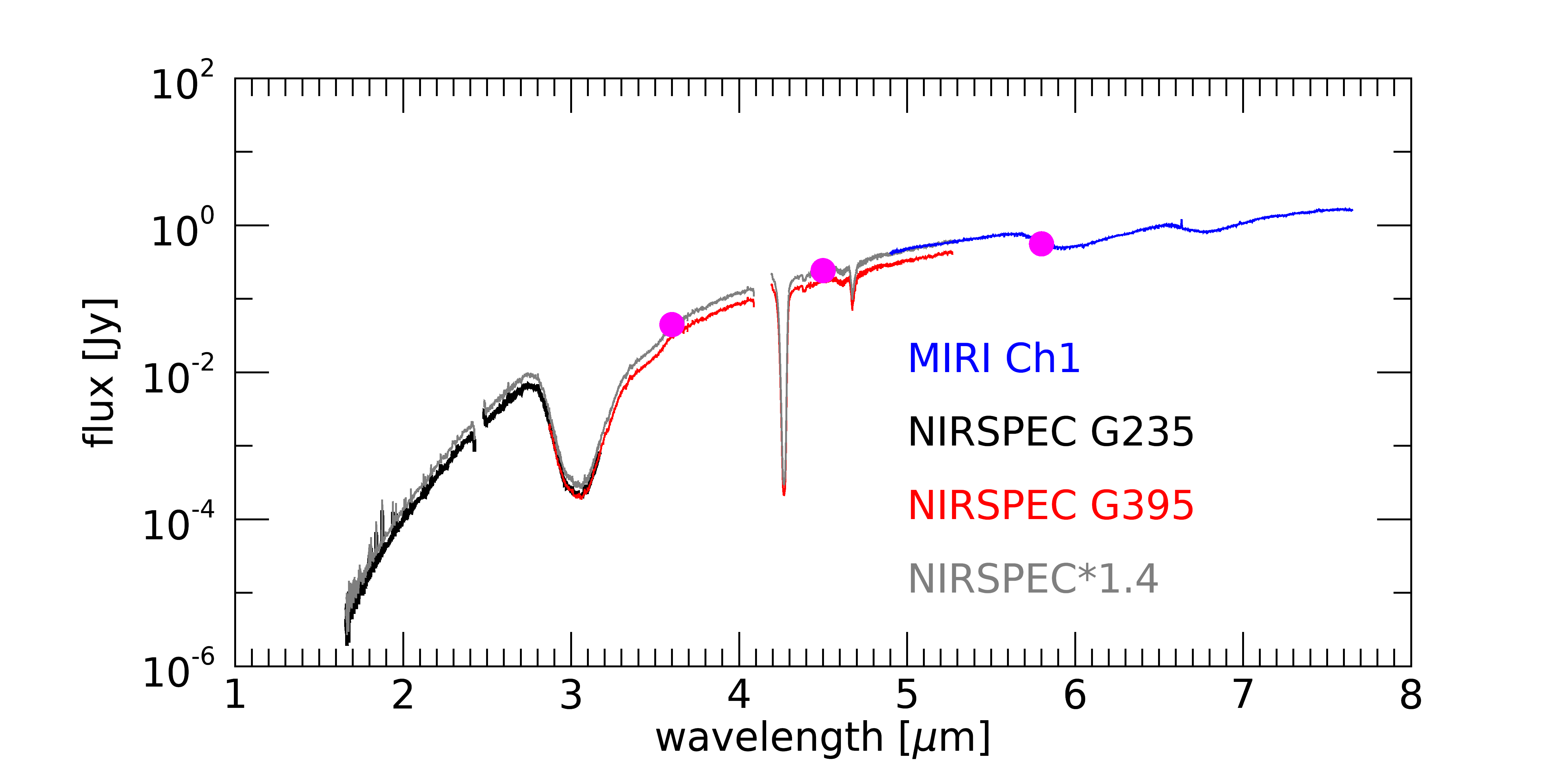}
    \caption{The upper panel shows the MIRI spectra in the individual sub-channels (black lines) and the final complete spectrum obtained by rescaling the individual sub-spectra to the Channel 1 SHORT (red line). The figure also displays the Spitzer IRS spectrum from \cite{noriega2004} (in blue) as well as archival photometric points from Spitzer IRAC and MIPS (circles), WISE (triangles) and AKARI (squares). The bottom panel displays the NIRSpec G235 (black) and G395 (red) spectra in comparison with the MIRI Channel 1 spectrum (blue). The NIRSpec spectrum rescaled to match MIRI and the IRAC photometric points is indicated in gray.  }
    \label{fig:calibration}
\end{figure*}

Figure \ref{fig:calibration} shows the MIRI (upper panel) and NIRSpec (lower panel) spectrum extracted at the HH46~IRS position, using an aperture proportional to 1.22$\lambda$/D. The figure plots the spectra of the individual sub-channels (MIRI) and gratings (NIRSpec) as well as the final spectrum obtained by rescaling them to the MIRI Channel 1 SHORT spectrum. The figure also reports the Spitzer spectrum of the source for comparison, as well as mid-IR photometry from different facilities. The difference between the Spitzer and JWST spectra might be due to the source variability or to different extraction procedures.

\section{Spectra of the jet and \htwo\ peak}

\begin{figure*}[ht]
    \centering

    \includegraphics[scale=0.7]{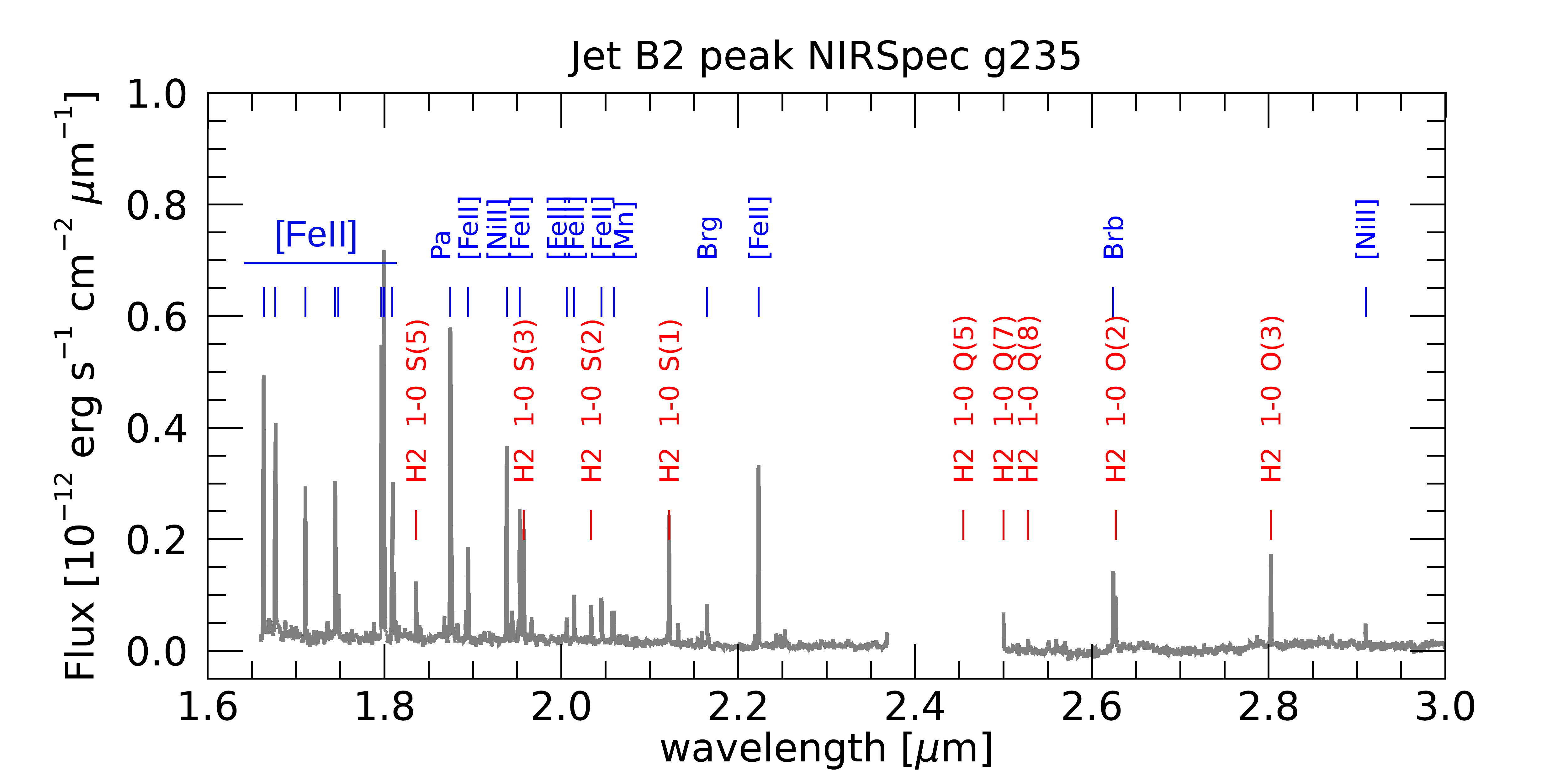}
    \includegraphics[scale=0.7]{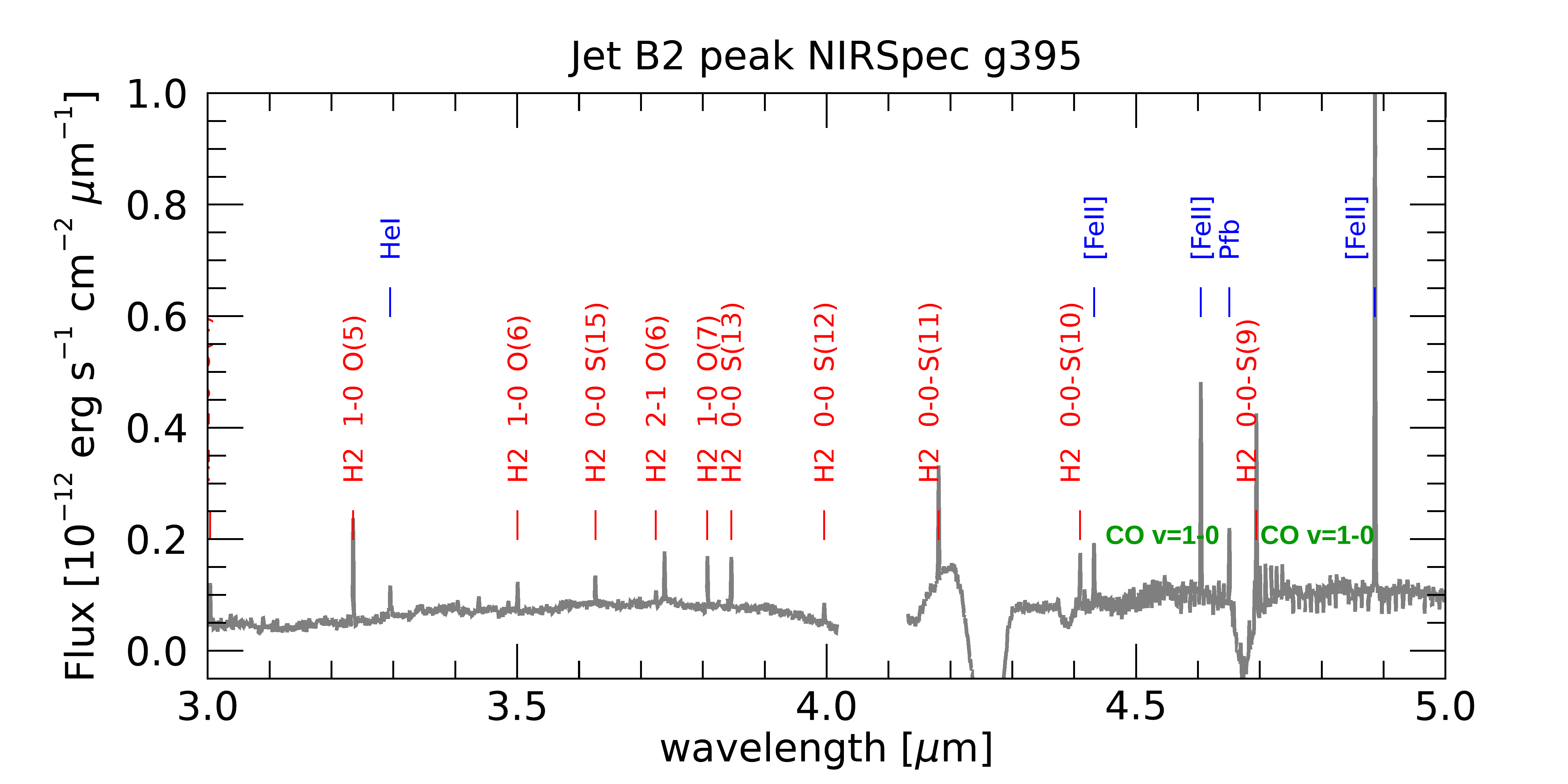}
    \includegraphics[scale=0.7]{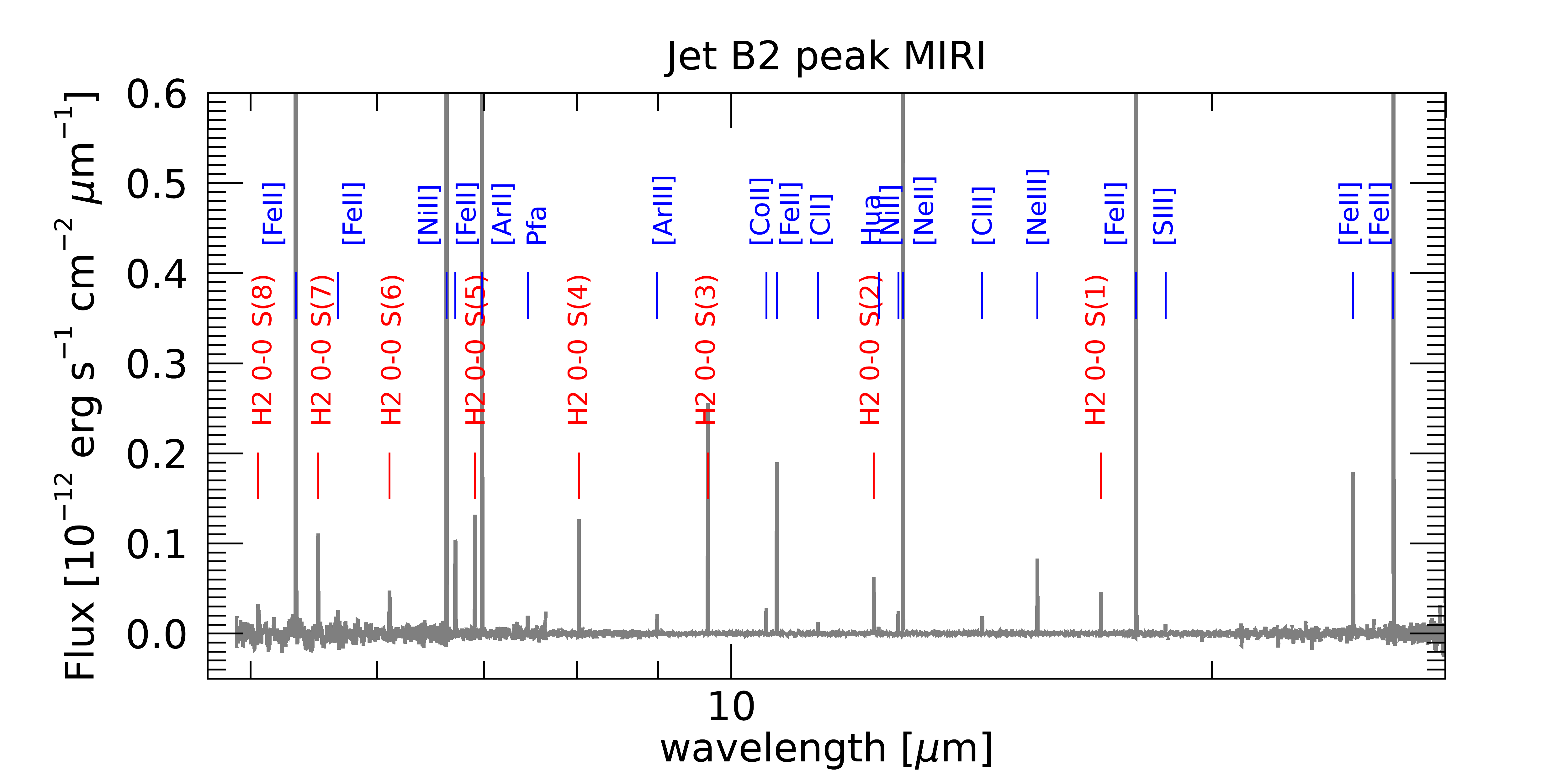}
    \caption{Complete NIRSpec and MIRI spectrum extracted from a circular aperture of 0\farcs 4 in radius centered on the B2 knot of the blue-shifted jet, with the major detected atomic (blue) and molecular (red for \htwo\ and green for CO) lines indicated. The MIRI spectrum has been continuum subtracted.}
    \label{fig:spettri_jet}
\end{figure*}

\begin{figure*}[ht]
    \centering

    \includegraphics[scale=0.7]{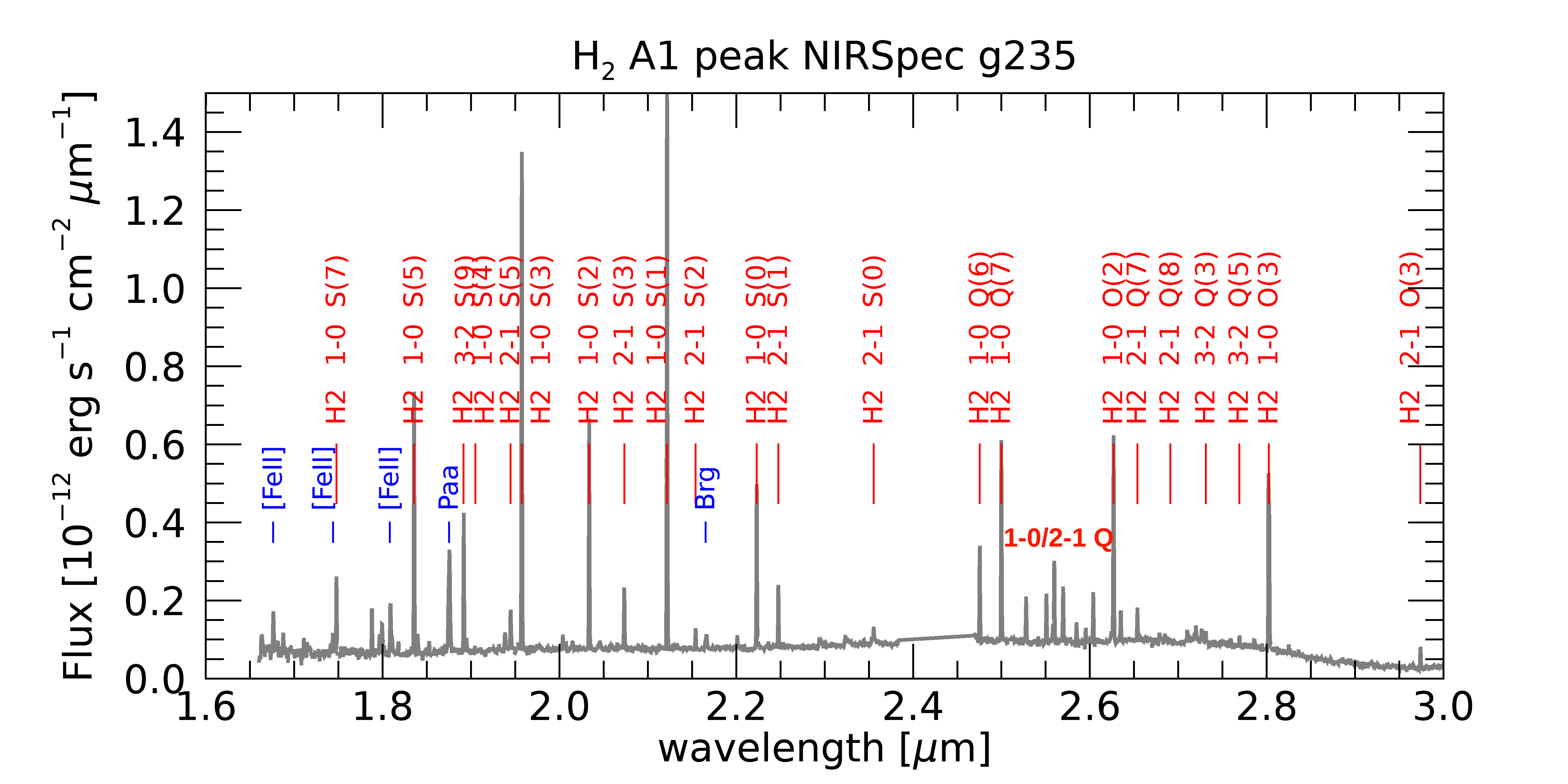}
    \includegraphics[scale=0.7]{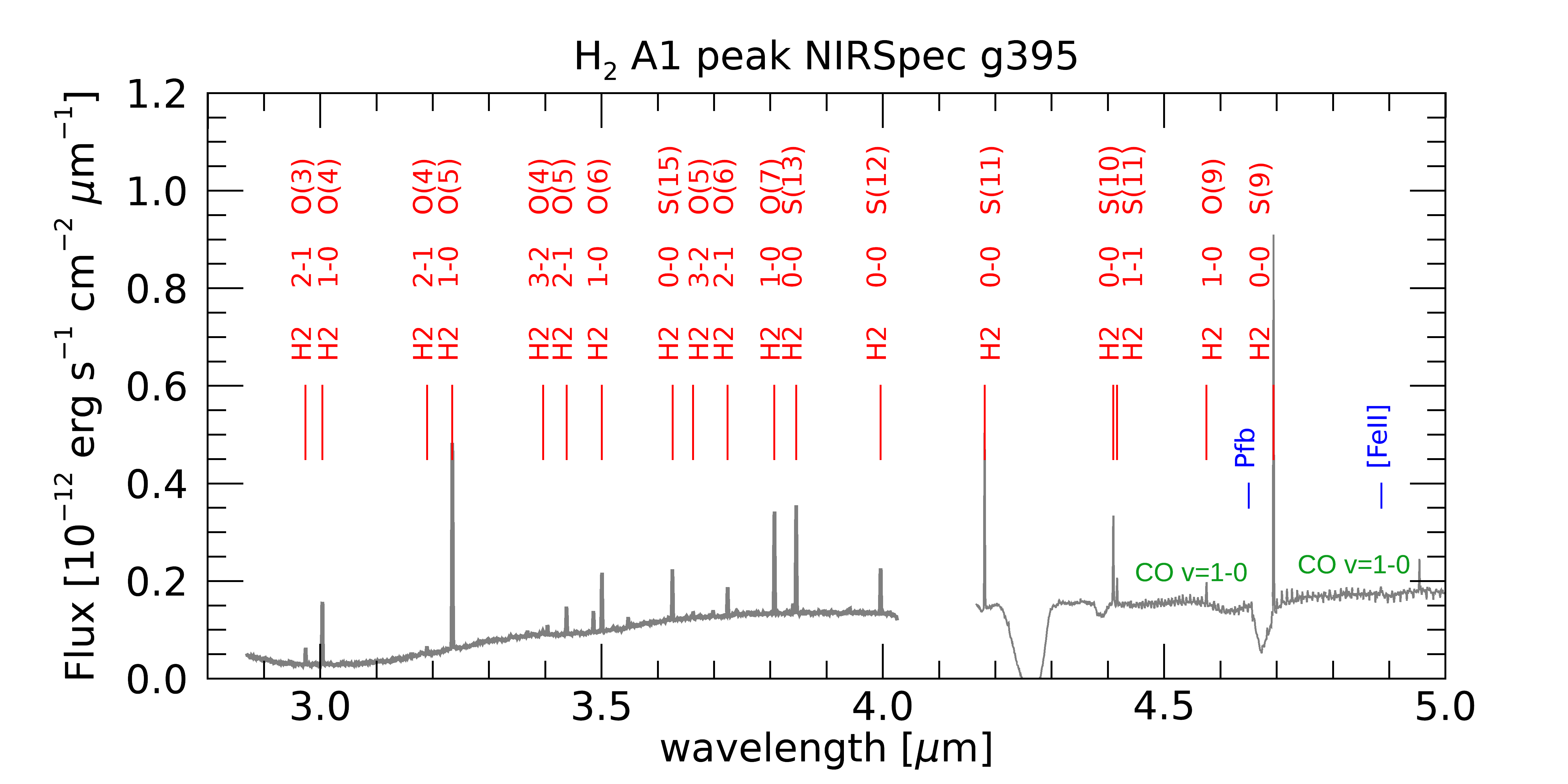}
    \includegraphics[scale=0.7]{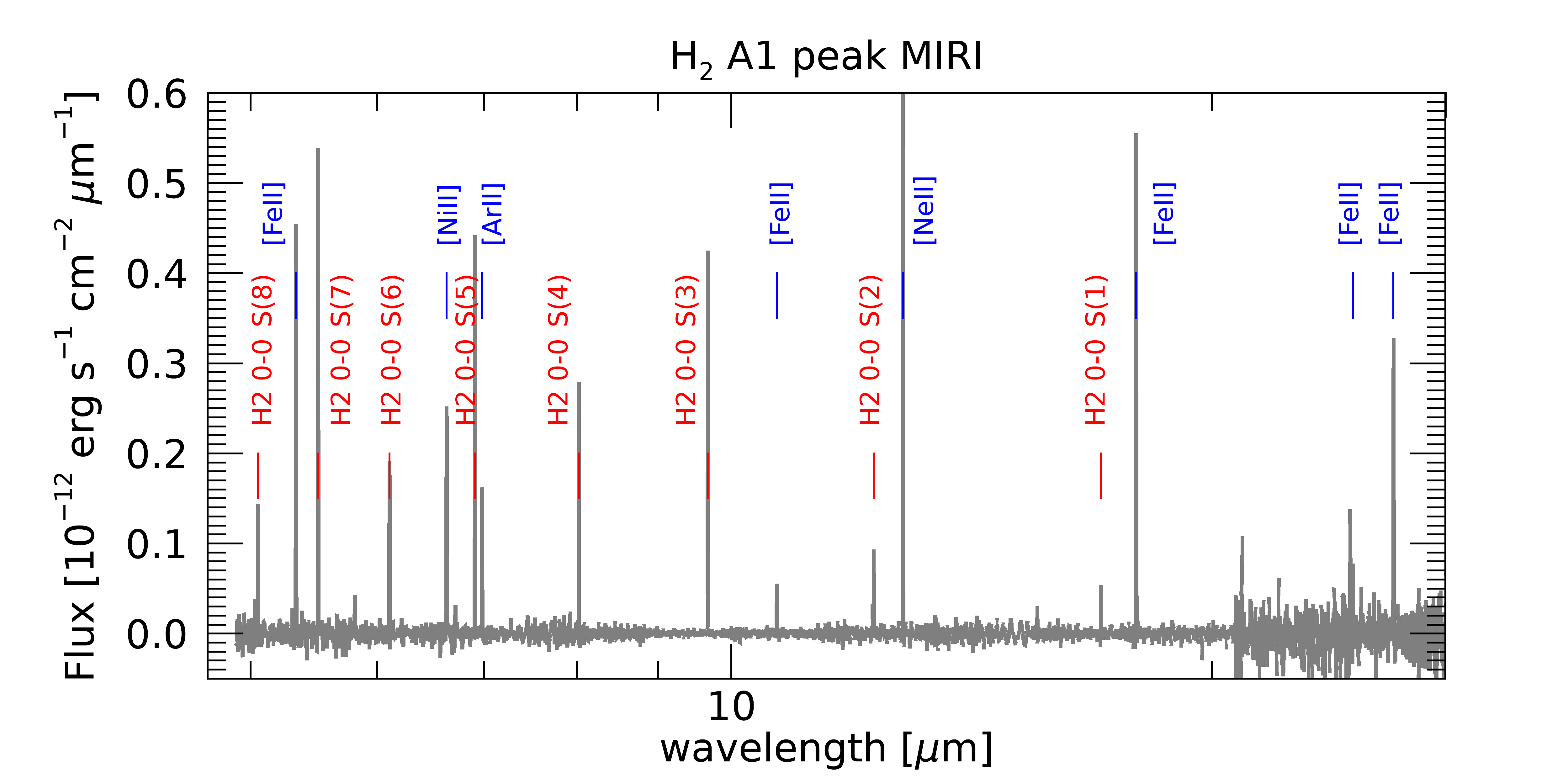}
    \caption{The same as Fig. \ref{fig:spettri_jet} but for an aperture centered at the peak of the \htwo\ knot A1. }
    \label{fig:spettri_H2}
\end{figure*}

Fig. \ref{fig:spettri_jet} and \ref{fig:spettri_H2} shows the NIRSpec and MIRI spectra extracted from a circular aperture of 0\farcs 4 in radius at the position of the B2 jet knot (08$^h$25$^m$43.88$^s$, $-$51\gradi 00\amin 34.55\asec ) and the \htwo\ A1 peak (08$^h$25$^m$43.86$^s$, $-$51\gradi 00\amin 35.69\asec ), respectively. These circular apertures are drawn in Fig. \ref{fig:MIRI_lines} and \ref{fig:MIRI_image}.

\section{Line list}

In Table \ref{table:atomic_lines} we list all the atomic lines detected in the blue-shifted jet, with their identification and relative intensity with respect to the \feii\ 5.3 \um\ line. Table \ref{table:h2_lines} summarises the \htwo\ transitions detected in the A1 knot, listed by vibrational level.

\include{Table1}
\include{Table2}

\bibliography{manuscript.bib}{}
\bibliographystyle{aasjournal}

\end{document}

%% file: Table1.tex
\begin{table*}[ht]
\caption{List of detected atomic lines in the jet}
\begin{tabular}{lllcccc}
\hline
\hline
Ion & I.P.$^b$ & Line ID & $\lambda$(vac) &  $T_{ex}^a$ & $I_\lambda/I_{[FeII]5.3\mu \rm m}$ & Grism/Chan\\\relax
& [eV] & & [\um]  & [K]  & &\\
\hline
 \feii & 7.90 & a$^4$D$_{1/2}$-a$^4$F$_{5/2}$ & 1.66422497 & 4083.22& 0.05163 & G235\\
& & a$^4$D$_{5/2}$-a$^4$F$_{7/2}$  & 1.8098952 &  11445.92& 0.06823& G235\\
& & a$^4$D$_{5/2}$-a$^4$F$_{3/2}$  & 1.8959259 &  12074.14& 0.025969& G235\\
& & a$^4$D$_{7/2}$-a$^4$F$_{5/2}$  & 1.9541410 &  11445.92& 0.03822& G235\\
& & a$^2$P$_{1/2}$-a$^4$P$_{1/2}$  & 2.0072612 &  27173.82& 0.007018& G235\\
& & a$^2$H$_{9/2}$-a$^2$G$_{9/2}$  & 2.0156760 & 29934.82& 0.013035& G235\\
& & a$^2$P$_{3/2}$-a$^4$P$_{5/2}$  & 2.0465833 & 26416.83& 0.01385& G235\\
& & a$^4$F$_{3/2}$-a$^6$D$_{3/2}$  & 4.4348339 & 4485.36& 0.00996& G395\\
& & a$^4$F$_{7/2}$-a$^6$D$_{7/2}$  & 4.8891395 & 3496.42 & 0.065055& G395\\\relax
& & a$^4$F$_{5/2}$-a$^6$D$_{3/2}$ & 5.0623456 & 4083.22&  0.01959 & ch1-SHORT\\
& & a$^4$F$_{9/2}$-a$^6$D$_{9/2}$ & 5.3401693 & 2694.25&  1.& ch1-SHORT\\
& &       a$^4$F$_{7/2}$-a$^6$D$_{5/2}$ & 5.6739070 & 3496.42& 0.01162& ch1-SHORT\\
& &       a$^4$F$_{9/2}$-a$^6$D$_{7/2}$ & 6.721277 & 2694.25&  0.04997& ch1-LONG\\
& &       a$^6$D$_{5/2}$-a$^6$D$_{9/2}$ & 14.977170 & 960.65 &0.001018 &ch3-MEDIUM \\
& &       a$^4$F$_{7/2}$-a$^4$F$_{9/2}$ & 17.92324 & 3496.42& 0.93672& ch3-LONG\\
& &       a$^4$F$_{5/2}$-a$^4$F$_{7/2}$ & 24.5019 & 4083.22& 0.254099& ch4-LONG\\
& &       a$^6$D$_{7/2}$-a$^6$D$_{9/2}$ & 25.988390 & 553.62$^*$ &0.90017& ch4-LONG \\\relax
[\ion{Ni}{2}] & 7.64 & $^2$F$_{7/2}$-$^4$F$_{9/2}$ & 1.93930 & 19495.97 &0.05259 &G235\\\relax
&               & $^2$F$_{7/2}$-$^4$F$_{5/2}$ & 2.91144 &  19495.97 & 0.005496& G235\\\relax
& & $^2$D$_{3/2}$-$^2$D$_{5/2}$ & 6.6360 & 2168.15   & 0.36327& ch1-LONG \\
& &         $^4$F$_{7/2}$-$^4$F$_{9/2}$ & 10.6822 & 13423.83    & 0.066096&ch2-LONG\\
& &       $^4$F$_{5/2}$-$^4$F$_{7/2}$ & 12.7288 & 14554.16  &0.01096 &ch3-SHORT\\\relax          
[\ion{Mn}{2}] & 7.43 & a$^5$D$_{4}$-a$^5$S$_{2}$   & 2.06063 & 20611.69 & 0.015145& G235\\\relax
[\ion{Ar}{2}] & 15.76 &  $^2$P$_{1/2}$-$^2$P$_{3/2}$ &  6.985274 & 2059.73$^*$ & 0.3737 & ch1-LONG\\\relax 
[\ion{Ar}{3}]& 27.63 & $^3$P$_{1}$-$^3$P$_{2}$ & 8.99138 & 1600.17$^*$ &  0.006188&ch2-MEDIUM\\\relax 
[\ion{Co}{2}] & 7.88 &a$^3$F$_{3}$-a$^3$F$_{4}$ & 10.522727 & 1367.30$^*$ &0.0093048& ch2-LONG\\\relax 
  &   &    a$^5$F$_{4}$-a$^5$F$_{5}$  & 14.7287 & 5796.81 & 0.002452&ch3-MEDIUM\\\relax
  &   &    a$^3$F$_{2}$-a$^3$F$_{3}$  & 15.45898 & 2298.01 & 0.002789 &ch3-MEDIUM\\\relax
[\ion{Cl}{1}] & 0 & $^2$P$_{1/2}$-$^2$P$_{3/2}$ & 11.333352 & 1269.51$^*$ &0.0044875 &ch2-LONG\\\relax 
[\ion{Cl}{2}]& 23.81 & $^3$P$_{1}$-$^3$P$_{2}$ & 14.3678 & 1001.39$^*$ &0.011506& ch3-MEDIUM\\\relax 
[\ion{Ne}{2}]& 21.56 & $^2$P$_{1/2}$-$^2$P$_{3/2}$ & 12.813548 & 1122.85$^*$ & 0.89572& ch3-SHORT\\\relax 
[\ion{Ne}{3}] & 40.96 &$^3$P$_{1}$-$^3$P$_{2}$ & 15.55505  & 924.96$^*$ &0.044233& ch3-LONG\\\relax 
[\ion{S}{1}] & 0 & $^3$P$_{1}$-$^3$P$_{2}$ & 25.2490 & 569.83$^*$ & 0.01844& ch4-LONG\\\relax
[\ion{S}{3}] & 23.24 & $^3$P$_{2}$-$^3$P$_{1}$ & 18.71303 & 1198.59 &0.008877& ch4-SHORT\\\relax
HI  & 13.598 & Pa$\alpha$ (4-3) & 1.875613 & 97492.31& 0.097326& G235\\
  & & Br$\gamma$ (7-4) & 2.166120& 107440.45 &0.007486 & G235\\
  &  & Br$\beta$ (6-4) & 2.625867& 106632.17 & 0.018529&  G235\\
  &  & Pf$\beta$ (7-5) & 4.653778& 107440.45& 0.008137&  G395\\
  & &Pf$\alpha$ (6-5) & 7.459858 & 153419.72 &  0.011506& ch1-LONG\\
  & & Hu$\alpha$ (7-6) & 12.371898 & 154582.66 & 0.004055& ch3-SHORT\\
\hline
\end{tabular}

$^a$ Excitation temperature of the upper level\\
$^b$ Ionisation potential of the $X^{i-1}$ ion \\
$^*$ Fundamental transition to the ground state\\
\label{table:atomic_lines}
\end{table*}

%% file: Table2.tex
\begin{table*}[ht]
\caption{Summary of the H$_2$ lines detected in the molecular peak}
\begin{tabular}{|l|l|c|}
\hline
\hline
v$_{u}$-v$_{l}$ & Rot. lines & $T_{ex}$[K] min-max $^a$ \\
\hline
0-0 & S(1)-S(15) & 354-1.5e4 \\
1-0 & S(0)-S(9), Q(1)-Q(13),O(2)-O(9) & 4160-1.1e4 \\
2-1 & S(0)-S(8),Q(1)-Q(8),O(3)-O(6) & 8087-1.4e4 \\
3-2 & S(1)/S(3)/S(9),Q(3)/Q(5),O(4)/O(5) & 1.2e4-1.3e4\\
\hline
\end{tabular}

$^a$ Minimum and maximum excitation temperature of the upper level\\
\label{table:h2_lines}
\end{table*}

%% file: manuscript.bbl
\begin{thebibliography}{}
\expandafter\ifx\csname natexlab\endcsname\relax\def\natexlab#1{#1}\fi
\providecommand{\url}[1]{\href{#1}{#1}}
\providecommand{\dodoi}[1]{doi:~\href{http://doi.org/#1}{\nolinkurl{#1}}}
\providecommand{\doeprint}[1]{\href{http://ascl.net/#1}{\nolinkurl{http://ascl.net/#1}}}
\providecommand{\doarXiv}[1]{\href{https://arxiv.org/abs/#1}{\nolinkurl{https://arxiv.org/abs/#1}}}

\bibitem[{{Agra-Amboage} {et~al.}(2014){Agra-Amboage}, {Cabrit}, {Dougados}, {Kristensen}, {Ibgui}, \& {Reunanen}}]{agra2014}
{Agra-Amboage}, V., {Cabrit}, S., {Dougados}, C., {et~al.} 2014, \aap, 564, A11, \dodoi{10.1051/0004-6361/201220488}

\bibitem[{{Alcal{\'a}} {et~al.}(2017){Alcal{\'a}}, {Manara}, {Natta}, {Frasca}, {Testi}, {Nisini}, {Stelzer}, {Williams}, {Antoniucci}, {Biazzo}, {Covino}, {Esposito}, {Getman}, \& {Rigliaco}}]{alcala2017}
{Alcal{\'a}}, J.~M., {Manara}, C.~F., {Natta}, A., {et~al.} 2017, \aap, 600, A20, \dodoi{10.1051/0004-6361/201629929}

\bibitem[{{Antoniucci} {et~al.}(2008){Antoniucci}, {Nisini}, {Giannini}, \& {Lorenzetti}}]{antoniucci2008}
{Antoniucci}, S., {Nisini}, B., {Giannini}, T., \& {Lorenzetti}, D. 2008, \aap, 479, 503, \dodoi{10.1051/0004-6361:20077468}

\bibitem[{{Antoniucci} {et~al.}(2017){Antoniucci}, {Nisini}, {Giannini}, {Rigliaco}, {Alcal{\'a}}, {Natta}, \& {Stelzer}}]{antoniucci2017}
{Antoniucci}, S., {Nisini}, B., {Giannini}, T., {et~al.} 2017, \aap, 599, A105, \dodoi{10.1051/0004-6361/201629683}

\bibitem[{{Arce} {et~al.}(2013){Arce}, {Mardones}, {Corder}, {Garay}, {Noriega-Crespo}, \& {Raga}}]{arce2013}
{Arce}, H.~G., {Mardones}, D., {Corder}, S.~A., {et~al.} 2013, \apj, 774, 39, \dodoi{10.1088/0004-637X/774/1/39}

\bibitem[{{Argyriou} {et~al.}(2023){Argyriou}, {Glasse}, {Law}, {Labiano}, {{\'A}lvarez-M{\'a}rquez}, {Patapis}, {Kavanagh}, {Gasman}, {Mueller}, {Larson}, {Vandenbussche}, {Glauser}, {Royer}, {Dicken}, {Harkett}, {Sargent}, {Engesser}, {Jones}, {Kendrew}, {Noriega-Crespo}, {Brandl}, {Rieke}, {Wright}, {Lee}, \& {Wells}}]{Argyriou2023}
{Argyriou}, I., {Glasse}, A., {Law}, D.~R., {et~al.} 2023, \aap, 675, A111, \dodoi{10.1051/0004-6361/202346489}

\bibitem[{{Bai} {et~al.}(2016){Bai}, {Ye}, {Goodman}, \& {Yuan}}]{bai2016}
{Bai}, X.-N., {Ye}, J., {Goodman}, J., \& {Yuan}, F. 2016, \apj, 818, 152, \dodoi{10.3847/0004-637X/818/2/152}

\bibitem[{{Bally}(2016)}]{bally2016}
{Bally}, J. 2016, \araa, 54, 491, \dodoi{10.1146/annurev-astro-081915-023341}

\bibitem[{{Banzatti} {et~al.}(2022){Banzatti}, {Abernathy}, {Brittain}, {Bosman}, {Pontoppidan}, {Boogert}, {Jensen}, {Carr}, {Najita}, {Grant}, {Sigler}, {Sanchez}, {Kern}, \& {Rayner}}]{banzatti2022}
{Banzatti}, A., {Abernathy}, K.~M., {Brittain}, S., {et~al.} 2022, \aj, 163, 174, \dodoi{10.3847/1538-3881/ac52f0}

\bibitem[{{Birney} {et~al.}(2023){Birney}, {Whelan}, {Dougados}, \& {Nisini}}]{birney2023}
{Birney}, M., {Whelan}, E., {Dougados}, C., \& {Nisini}, B. 2023, \aap, submitted

\bibitem[{{B{\"o}ker} {et~al.}(2022){B{\"o}ker}, {Arribas}, {L{\"u}tzgendorf}, {Alves de Oliveira}, {Beck}, {Birkmann}, {Bunker}, {Charlot}, {de Marchi}, {Ferruit}, {Giardino}, {Jakobsen}, {Kumari}, {L{\'o}pez-Caniego}, {Maiolino}, {Manjavacas}, {Marston}, {Moseley}, {Muzerolle}, {Ogle}, {Pirzkal}, {Rauscher}, {Rawle}, {Rix}, {Sabbi}, {Sargent}, {Sirianni}, {te Plate}, {Valenti}, {Willott}, \& {Zeidler}}]{boker2022}
{B{\"o}ker}, T., {Arribas}, S., {L{\"u}tzgendorf}, N., {et~al.} 2022, \aap, 661, A82, \dodoi{10.1051/0004-6361/202142589}

\bibitem[{{Boogert} {et~al.}(2004){Boogert}, {Pontoppidan}, {Lahuis}, {J{\o}rgensen}, {Augereau}, {Blake}, {Brooke}, {Brown}, {Dullemond}, {Evans}, {Geers}, {Hogerheijde}, {Kessler-Silacci}, {Knez}, {Morris}, {Noriega-Crespo}, {Sch{\"o}ier}, {van Dishoeck}, {Allen}, {Harvey}, {Koerner}, {Mundy}, {Myers}, {Padgett}, {Sargent}, \& {Stapelfeldt}}]{boogert2004}
{Boogert}, A.~C.~A., {Pontoppidan}, K.~M., {Lahuis}, F., {et~al.} 2004, \apjs, 154, 359, \dodoi{10.1086/422556}

\bibitem[{{Bushouse} {et~al.}(2023){Bushouse}, {Eisenhamer}, {Dencheva}, {Davies}, {Greenfield}, {Morrison}, {Hodge}, {Simon}, {Grumm}, {Droettboom}, {Slavich}, {Sosey}, {Pauly}, {Miller}, {Jedrzejewski}, {Hack}, {Davis}, {Crawford}, {Law}, {Gordon}, {Regan}, {Cara}, {MacDonald}, {Bradley}, {Shanahan}, {Jamieson}, {Teodoro}, \& {Williams}}]{bushouse2023}
{Bushouse}, H., {Eisenhamer}, J., {Dencheva}, N., {et~al.} 2023, {JWST Calibration Pipeline}, 1.9.6, Zenodo,  Zenodo, \dodoi{10.5281/zenodo.7714020}

\bibitem[{{de Valon} {et~al.}(2022){de Valon}, {Dougados}, {Cabrit}, {Louvet}, {Zapata}, \& {Mardones}}]{devalon2022}
{de Valon}, A., {Dougados}, C., {Cabrit}, S., {et~al.} 2022, \aap, 668, A78, \dodoi{10.1051/0004-6361/202141316}

\bibitem[{{Dionatos} {et~al.}(2010){Dionatos}, {Nisini}, {Cabrit}, {Kristensen}, \& {Pineau Des For{\^e}ts}}]{dionatos2010}
{Dionatos}, O., {Nisini}, B., {Cabrit}, S., {Kristensen}, L., \& {Pineau Des For{\^e}ts}, G. 2010, \aap, 521, A7, \dodoi{10.1051/0004-6361/200913650}

\bibitem[{{Eisl\"offel} {et~al.}(1994){Eisl\"offel}, {Davis}, {Ray}, \& {Mundt}}]{eisloffelh21994}
{Eisl\"offel}, J., {Davis}, C.~J., {Ray}, T.~P., \& {Mundt}, R. 1994, \apjl, 422, L91, \dodoi{10.1086/187220}

\bibitem[{{Eisl\"offel} \& {Mundt}(1994)}]{eisloffelpm1994}
{Eisl\"offel}, J., \& {Mundt}, R. 1994, \aap, 284, 530

\bibitem[{{Erkal} {et~al.}(2021){Erkal}, {Nisini}, {Coffey}, {Bacciotti}, {Hartigan}, {Antoniucci}, {Giannini}, {Eisl{\"o}ffel}, \& {Manara}}]{erkal2021}
{Erkal}, J., {Nisini}, B., {Coffey}, D., {et~al.} 2021, \apj, 919, 23, \dodoi{10.3847/1538-4357/ac06c5}

\bibitem[{{Federman} {et~al.}(2023){Federman}, {Megeath}, {Rubinstein}, {Gutermuth}, {Narang}, {Tyagi}, {Manoj}, {Anglada}, {Atnagulov}, {Beuther}, {Bourke}, {Brunken}, {Garatti}, {Evans}, {Fischer}, {Furlan}, {Green}, {Habel}, {Hartmann}, {Karnath}, {Klaassen}, {Linz}, {Looney}, {Osorio}, {Muzerolle Page}, {Pokhrel}, {Rahatgaonkar}, {Rocha}, {Sheehan}, {Slavicinska}, {Stanke}, {Stutz}, {Tobin}, {Tychoniec}, {Van Dishoeck}, {Watson}, {Wolk}, \& {Yang}}]{federman2023}
{Federman}, S., {Megeath}, S.~T., {Rubinstein}, A.~E., {et~al.} 2023, arXiv e-prints, arXiv:2310.03803, \dodoi{10.48550/arXiv.2310.03803}

\bibitem[{{Ferreira}(1997)}]{ferreira1997}
{Ferreira}, J. 1997, \aap, 319, 340, \dodoi{10.48550/arXiv.astro-ph/9607057}

\bibitem[{{Frank} {et~al.}(2014){Frank}, {Ray}, {Cabrit}, {Hartigan}, {Arce}, {Bacciotti}, {Bally}, {Benisty}, {Eisl{\"o}ffel}, {G{\"u}del}, {Lebedev}, {Nisini}, \& {Raga}}]{frank2014}
{Frank}, A., {Ray}, T.~P., {Cabrit}, S., {et~al.} 2014, in Protostars and Planets VI, ed. H.~{Beuther}, R.~S. {Klessen}, C.~P. {Dullemond}, \& T.~{Henning}, 451--474, \dodoi{10.2458/azu_uapress_9780816531240-ch020}

\bibitem[{{Garcia Lopez} {et~al.}(2010){Garcia Lopez}, {Nisini}, {Eisl{\"o}ffel}, {Giannini}, {Bacciotti}, \& {Podio}}]{garcialopez2010}
{Garcia Lopez}, R., {Nisini}, B., {Eisl{\"o}ffel}, J., {et~al.} 2010, \aap, 511, A5, \dodoi{10.1051/0004-6361/200913304}

\bibitem[{{Gasman} {et~al.}(2023){Gasman}, {Argyriou}, {Sloan}, {Aringer}, {{\'A}lvarez-M{\'a}rquez}, {Fox}, {Glasse}, {Glauser}, {Jones}, {Justtanont}, {Kavanagh}, {Klaassen}, {Labiano}, {Larson}, {Law}, {Mueller}, {Nayak}, {Noriega-Crespo}, {Patapis}, {Royer}, \& {Vandenbussche}}]{gasman2023}
{Gasman}, D., {Argyriou}, I., {Sloan}, G.~C., {et~al.} 2023, \aap, 673, A102, \dodoi{10.1051/0004-6361/202245633}

\bibitem[{{Giannini} {et~al.}(2011){Giannini}, {Nisini}, {Neufeld}, {Yuan}, {Antoniucci}, \& {Gusdorf}}]{giannini2011}
{Giannini}, T., {Nisini}, B., {Neufeld}, D., {et~al.} 2011, \apj, 738, 80, \dodoi{10.1088/0004-637X/738/1/80}

\bibitem[{{Glassgold} {et~al.}(2007){Glassgold}, {Najita}, \& {Igea}}]{glassgold2007}
{Glassgold}, A.~E., {Najita}, J.~R., \& {Igea}, J. 2007, \apj, 656, 515, \dodoi{10.1086/510013}

\bibitem[{{Grant} {et~al.}(2023){Grant}, {van Dishoeck}, {Tabone}, {Gasman}, {Henning}, {Kamp}, {G{\"u}del}, {Lagage}, {Bettoni}, {Perotti}, {Christiaens}, {Samland}, {Arabhavi}, {Argyriou}, {Abergel}, {Absil}, {Barrado}, {Boccaletti}, {Bouwman}, {o Garatti}, {Geers}, {Glauser}, {Guadarrama}, {Jang}, {Kanwar}, {Lahuis}, {Morales-Calder{\'o}n}, {Mueller}, {Nehm{\'e}}, {Olofsson}, {Pantin}, {Pawellek}, {Ray}, {Rodgers-Lee}, {Scheithauer}, {Schreiber}, {Schwarz}, {Temmink}, {Vandenbussche}, {Vlasblom}, {Waters}, {Wright}, {Colina}, {Greve}, {Justannont}, \& {{\"O}stlin}}]{grant2023}
{Grant}, S.~L., {van Dishoeck}, E.~F., {Tabone}, B., {et~al.} 2023, \apjl, 947, L6, \dodoi{10.3847/2041-8213/acc44b}

\bibitem[{{Greene} {et~al.}(2018){Greene}, {Gully-Santiago}, \& {Barsony}}]{greene2018}
{Greene}, T.~P., {Gully-Santiago}, M.~A., \& {Barsony}, M. 2018, \apj, 862, 85, \dodoi{10.3847/1538-4357/aacc6c}

\bibitem[{{Hartigan} {et~al.}(2005){Hartigan}, {Heathcote}, {Morse}, {Reipurth}, \& {Bally}}]{hartigan2005}
{Hartigan}, P., {Heathcote}, S., {Morse}, J.~A., {Reipurth}, B., \& {Bally}, J. 2005, \aj, 130, 2197, \dodoi{10.1086/491673}

\bibitem[{{Hartigan} {et~al.}(1987){Hartigan}, {Raymond}, \& {Hartmann}}]{hartigan1987}
{Hartigan}, P., {Raymond}, J., \& {Hartmann}, L. 1987, \apj, 316, 323, \dodoi{10.1086/165204}

\bibitem[{{Hartigan} {et~al.}(2011){Hartigan}, {Frank}, {Foster}, {Wilde}, {Douglas}, {Rosen}, {Coker}, {Blue}, \& {Hansen}}]{hartigan2011}
{Hartigan}, P., {Frank}, A., {Foster}, J.~M., {et~al.} 2011, \apj, 736, 29, \dodoi{10.1088/0004-637X/736/1/29}

\bibitem[{{Heathcote} {et~al.}(1996){Heathcote}, {Morse}, {Hartigan}, {Reipurth}, {Schwartz}, {Bally}, \& {Stone}}]{heathcote1996}
{Heathcote}, S., {Morse}, J.~A., {Hartigan}, P., {et~al.} 1996, \aj, 112, 1141, \dodoi{10.1086/118085}

\bibitem[{{Helmich} {et~al.}(1996){Helmich}, {van Dishoeck}, {Black}, {de Graauw}, {Beintema}, {Heras}, {Lahuis}, {Morris}, \& {Valentijn}}]{helmich1996}
{Helmich}, F.~P., {van Dishoeck}, E.~F., {Black}, J.~H., {et~al.} 1996, \aap, 315, L173

\bibitem[{{Hollenbach} \& {Gorti}(2009)}]{hollenbach2009}
{Hollenbach}, D., \& {Gorti}, U. 2009, \apj, 703, 1203, \dodoi{10.1088/0004-637X/703/2/1203}

\bibitem[{{Hollenbach} \& {McKee}(1989)}]{hollenbach1989}
{Hollenbach}, D., \& {McKee}, C.~F. 1989, \apj, 342, 306, \dodoi{10.1086/167595}

\bibitem[{{Jakobsen} {et~al.}(2022){Jakobsen}, {Ferruit}, {Alves de Oliveira}, {Arribas}, {Bagnasco}, {Barho}, {Beck}, {Birkmann}, {B{\"o}ker}, {Bunker}, {Charlot}, {de Jong}, {de Marchi}, {Ehrenwinkler}, {Falcolini}, {Fels}, {Franx}, {Franz}, {Funke}, {Giardino}, {Gnata}, {Holota}, {Honnen}, {Jensen}, {Jentsch}, {Johnson}, {Jollet}, {Karl}, {Kling}, {K{\"o}hler}, {Kolm}, {Kumari}, {Lander}, {Lemke}, {L{\'o}pez-Caniego}, {L{\"u}tzgendorf}, {Maiolino}, {Manjavacas}, {Marston}, {Maschmann}, {Maurer}, {Messerschmidt}, {Moseley}, {Mosner}, {Mott}, {Muzerolle}, {Pirzkal}, {Pittet}, {Plitzke}, {Posselt}, {Rapp}, {Rauscher}, {Rawle}, {Rix}, {R{\"o}del}, {Rumler}, {Sabbi}, {Salvignol}, {Schmid}, {Sirianni}, {Smith}, {Strada}, {te Plate}, {Valenti}, {Wettemann}, {Wiehe}, {Wiesmayer}, {Willott}, {Wright}, {Zeidler}, \& {Zincke}}]{jakobsen2022}
{Jakobsen}, P., {Ferruit}, P., {Alves de Oliveira}, C., {et~al.} 2022, \aap, 661, A80, \dodoi{10.1051/0004-6361/202142663}

\bibitem[{{Lee}(2020)}]{lee2020}
{Lee}, C.-F. 2020, \aapr, 28, 1, \dodoi{10.1007/s00159-020-0123-7}

\bibitem[{{Liu} {et~al.}(2016){Liu}, {Shang}, {Herczeg}, \& {Walter}}]{liu2016}
{Liu}, C.-F., {Shang}, H., {Herczeg}, G.~J., \& {Walter}, F.~M. 2016, \apj, 832, 153, \dodoi{10.3847/0004-637X/832/2/153}

\bibitem[{{Masciadri} \& {Raga}(2002)}]{masciadri2002}
{Masciadri}, E., \& {Raga}, A.~C. 2002, \apj, 568, 733, \dodoi{10.1086/338767}

\bibitem[{{Maurri} {et~al.}(2014){Maurri}, {Bacciotti}, {Podio}, {Eisl{\"o}ffel}, {Ray}, {Mundt}, {Locatelli}, \& {Coffey}}]{maurri2014}
{Maurri}, L., {Bacciotti}, F., {Podio}, L., {et~al.} 2014, \aap, 565, A110, \dodoi{10.1051/0004-6361/201117510}

\bibitem[{{McClure} {et~al.}(2023){McClure}, {Rocha}, {Pontoppidan}, {Crouzet}, {Chu}, {Dartois}, {Lamberts}, {Noble}, {Pendleton}, {Perotti}, {Qasim}, {Rachid}, {Smith}, {Sun}, {Beck}, {Boogert}, {Brown}, {Caselli}, {Charnley}, {Cuppen}, {Dickinson}, {Drozdovskaya}, {Egami}, {Erkal}, {Fraser}, {Garrod}, {Harsono}, {Ioppolo}, {Jim{\'e}nez-Serra}, {Jin}, {J{\o}rgensen}, {Kristensen}, {Lis}, {McCoustra}, {McGuire}, {Melnick}, {{\~A}-berg}, {Palumbo}, {Shimonishi}, {Sturm}, {van Dishoeck}, \& {Linnartz}}]{mcclure2023}
{McClure}, M.~K., {Rocha}, W.~R.~M., {Pontoppidan}, K.~M., {et~al.} 2023, Nature Astronomy, 7, 431, \dodoi{10.1038/s41550-022-01875-w}

\bibitem[{{Nisini}(2003)}]{nisini2003}
{Nisini}, B. 2003, \apss, 287, 207, \dodoi{10.1023/B:ASTR.0000006225.10230.dd}

\bibitem[{{Nisini} {et~al.}(2005){Nisini}, {Antoniucci}, {Giannini}, \& {Lorenzetti}}]{nisini2005}
{Nisini}, B., {Antoniucci}, S., {Giannini}, T., \& {Lorenzetti}, D. 2005, \aap, 429, 543, \dodoi{10.1051/0004-6361:20041409}

\bibitem[{{Nisini} {et~al.}(2016){Nisini}, {Giannini}, {Antoniucci}, {Alcal{\'a}}, {Bacciotti}, \& {Podio}}]{nisini2016}
{Nisini}, B., {Giannini}, T., {Antoniucci}, S., {et~al.} 2016, \aap, 595, A76, \dodoi{10.1051/0004-6361/201628853}

\bibitem[{{Nisini} {et~al.}(2015){Nisini}, {Santangelo}, {Giannini}, {Antoniucci}, {Cabrit}, {Codella}, {Davis}, {Eisl{\"o}ffel}, {Kristensen}, {Herczeg}, {Neufeld}, \& {van Dishoeck}}]{nisini2015}
{Nisini}, B., {Santangelo}, G., {Giannini}, T., {et~al.} 2015, \apj, 801, 121, \dodoi{10.1088/0004-637X/801/2/121}

\bibitem[{{Noriega-Crespo} {et~al.}(2004){Noriega-Crespo}, {Morris}, {Marleau}, {Carey}, {Boogert}, {van Dishoeck}, {Evans}, {Keene}, {Muzerolle}, {Stapelfeldt}, {Pontoppidan}, {Lowrance}, {Allen}, \& {Bourke}}]{noriega2004}
{Noriega-Crespo}, A., {Morris}, P., {Marleau}, F.~R., {et~al.} 2004, \apjs, 154, 352, \dodoi{10.1086/422819}

\bibitem[{{Pascucci} {et~al.}(2023){Pascucci}, {Cabrit}, {Edwards}, {Gorti}, {Gressel}, \& {Suzuki}}]{pascucci2023}
{Pascucci}, I., {Cabrit}, S., {Edwards}, S., {et~al.} 2023, in Astronomical Society of the Pacific Conference Series, Vol. 534, Protostars and Planets VII, ed. S.~{Inutsuka}, Y.~{Aikawa}, T.~{Muto}, K.~{Tomida}, \& M.~{Tamura}, 567, \dodoi{10.48550/arXiv.2203.10068}

\bibitem[{{Patapis} {et~al.}(2023){Patapis}, {Argyriou}, {Law}, {Glauser}, {Glasse}, {Labiano}, {{\'A}lvarez-M{\'a}rquez}, {Kavanagh}, {Gasman}, {Mueller}, {Larson}, {Vandenbussche}, {Klaassen}, {Guillard}, \& {Wright}}]{patapis2023}
{Patapis}, P., {Argyriou}, I., {Law}, D.~R., {et~al.} 2023, arXiv e-prints, arXiv:2307.01025, \dodoi{10.48550/arXiv.2307.01025}

\bibitem[{{Pelletier} \& {Pudritz}(1992)}]{pelletier1992}
{Pelletier}, G., \& {Pudritz}, R.~E. 1992, \apj, 394, 117, \dodoi{10.1086/171565}

\bibitem[{{Perna} {et~al.}(2023){Perna}, {Arribas}, {Marshall}, {D'Eugenio}, {{\"U}bler}, {Bunker}, {Charlot}, {Carniani}, {Jakobsen}, {Maiolino}, {Rodr{\'\i}guez Del Pino}, {Willott}, {B{\"o}ker}, {Circosta}, {Cresci}, {Curti}, {Husemann}, {Kumari}, {Lamperti}, {P{\'e}rez-Gonz{\'a}lez}, \& {Scholtz}}]{perna2023}
{Perna}, M., {Arribas}, S., {Marshall}, M., {et~al.} 2023, arXiv e-prints, arXiv:2304.06756, \dodoi{10.48550/arXiv.2304.06756}

\bibitem[{{Podio} {et~al.}(2021){Podio}, {Tabone}, {Codella}, {Gueth}, {Maury}, {Cabrit}, {Lefloch}, {Maret}, {Belloche}, {Andr{\'e}}, {Anderl}, {Gaudel}, \& {Testi}}]{podio2021}
{Podio}, L., {Tabone}, B., {Codella}, C., {et~al.} 2021, \aap, 648, A45, \dodoi{10.1051/0004-6361/202038429}

\bibitem[{{Rabenanahary} {et~al.}(2022){Rabenanahary}, {Cabrit}, {Meliani}, \& {Pineau des For{\^e}ts}}]{Rabenanahary2022}
{Rabenanahary}, M., {Cabrit}, S., {Meliani}, Z., \& {Pineau des For{\^e}ts}, G. 2022, \aap, 664, A118, \dodoi{10.1051/0004-6361/202243139}

\bibitem[{{Ray} {et~al.}(2023){Ray}, {McCaughrean}, {Caratti o Garatti}, {Kavanagh}, {Justtanont}, {van Dishoeck}, {Reitsma}, {Beuther}, {Francis}, {Gieser}, {Klaassen}, {Perotti}, {Tychoniec}, {van Gelder}, {Colina}, {Greve}, {G{\"u}del}, {Henning}, {Lagage}, {{\"O}stlin}, {Vandenbussche}, {Waelkens}, \& {Wright}}]{ray2023}
{Ray}, T.~P., {McCaughrean}, M.~J., {Caratti o Garatti}, A., {et~al.} 2023, \nat, 622, 48, \dodoi{10.1038/s41586-023-06551-1}

\bibitem[{{Reipurth} {et~al.}(2000){Reipurth}, {Yu}, {Heathcote}, {Bally}, \& {Rodr{\'\i}guez}}]{reipurth2000}
{Reipurth}, B., {Yu}, K.~C., {Heathcote}, S., {Bally}, J., \& {Rodr{\'\i}guez}, L.~F. 2000, \aj, 120, 1449, \dodoi{10.1086/301510}

\bibitem[{{Rieke} {et~al.}(2015){Rieke}, {Wright}, {B{\"o}ker}, {Bouwman}, {Colina}, {Glasse}, {Gordon}, {Greene}, {G{\"u}del}, {Henning}, {Justtanont}, {Lagage}, {Meixner}, {N{\o}rgaard-Nielsen}, {Ray}, {Ressler}, {van Dishoeck}, \& {Waelkens}}]{rieke2015}
{Rieke}, G.~H., {Wright}, G.~S., {B{\"o}ker}, T., {et~al.} 2015, \pasp, 127, 584, \dodoi{10.1086/682252}

\bibitem[{{Rigliaco} {et~al.}(2015){Rigliaco}, {Pascucci}, {Duchene}, {Edwards}, {Ardila}, {Grady}, {Mendigut{\'\i}a}, {Montesinos}, {Mulders}, {Najita}, {Carpenter}, {Furlan}, {Gorti}, {Meijerink}, \& {Meyer}}]{rigliaco2015}
{Rigliaco}, E., {Pascucci}, I., {Duchene}, G., {et~al.} 2015, \apj, 801, 31, \dodoi{10.1088/0004-637X/801/1/31}

\bibitem[{{Rocha} {et~al.}(2023){Rocha}, {van Dishoeck}, {Ressler}, {van Gelder}, {Slavicinska}, {Brunken}, {Linnartz}, {Ray}, {Beuther}, {Garatti}, {Geers}, {Kavanagh}, {Klaassen}, {Justannont}, {Chen}, {Francis}, {Gieser}, {Perotti}, {Tychoniec}, {Barsony}, {Majumdar}, {le Gouellec}, {Chu}, {Lew}, {Henning}, \& {Wright}}]{rocha2023}
{Rocha}, W.~R.~M., {van Dishoeck}, E.~F., {Ressler}, M.~E., {et~al.} 2023, arXiv e-prints, arXiv:2312.06834, \dodoi{10.48550/arXiv.2312.06834}

\bibitem[{{Salyk} {et~al.}(2013){Salyk}, {Herczeg}, {Brown}, {Blake}, {Pontoppidan}, \& {van Dishoeck}}]{salyk2013}
{Salyk}, C., {Herczeg}, G.~J., {Brown}, J.~M., {et~al.} 2013, \apj, 769, 21, \dodoi{10.1088/0004-637X/769/1/21}

\bibitem[{{Shang} {et~al.}(2020){Shang}, {Krasnopolsky}, {Liu}, \& {Wang}}]{shang2020}
{Shang}, H., {Krasnopolsky}, R., {Liu}, C.-F., \& {Wang}, L.-Y. 2020, \apj, 905, 116, \dodoi{10.3847/1538-4357/abbdb0}

\bibitem[{{Shang} {et~al.}(2023){Shang}, {Liu}, {Krasnopolsky}, \& {Wang}}]{shang2023}
{Shang}, H., {Liu}, C.-F., {Krasnopolsky}, R., \& {Wang}, L.-Y. 2023, \apj, 944, 230, \dodoi{10.3847/1538-4357/aca763}

\bibitem[{{Tabone} {et~al.}(2023){Tabone}, {Bettoni}, {van Dishoeck}, {Arabhavi}, {Grant}, {Gasman}, {Henning}, {Kamp}, {G{\"u}del}, {Lagage}, {Ray}, {Vandenbussche}, {Abergel}, {Absil}, {Argyriou}, {Barrado}, {Boccaletti}, {Bouwman}, {Caratti o Garatti}, {Geers}, {Glauser}, {Justannont}, {Lahuis}, {Mueller}, {Nehm{\'e}}, {Olofsson}, {Pantin}, {Scheithauer}, {Waelkens}, {Waters}, {Black}, {Christiaens}, {Guadarrama}, {Morales-Calder{\'o}n}, {Jang}, {Kanwar}, {Pawellek}, {Perotti}, {Perrin}, {Rodgers-Lee}, {Samland}, {Schreiber}, {Schwarz}, {Colina}, {{\"O}stlin}, \& {Wright}}]{tabone2023}
{Tabone}, B., {Bettoni}, G., {van Dishoeck}, E.~F., {et~al.} 2023, Nature Astronomy, 7, 805, \dodoi{10.1038/s41550-023-01965-3}

\bibitem[{{Urso} {et~al.}(2022){Urso}, {H{\'e}nault}, {Brunetto}, {Baklouti}, {Baratta}, {Djouadi}, {Elsaesser}, {Scir{\`e}}, {Strazzulla}, \& {Palumbo}}]{urso2022}
{Urso}, R.~G., {H{\'e}nault}, E., {Brunetto}, R., {et~al.} 2022, \aap, 668, A169, \dodoi{10.1051/0004-6361/202244522}

\bibitem[{{van Dokkum}(2001)}]{vandokkum2001}
{van Dokkum}, P.~G. 2001, \pasp, 113, 1420, \dodoi{10.1086/323894}

\bibitem[{{van Kempen} {et~al.}(2009){van Kempen}, {van Dishoeck}, {G{\"u}sten}, {Kristensen}, {Schilke}, {Hogerheijde}, {Boland}, {Nefs}, {Menten}, {Baryshev}, \& {Wyrowski}}]{vankempen2009}
{van Kempen}, T.~A., {van Dishoeck}, E.~F., {G{\"u}sten}, R., {et~al.} 2009, \aap, 501, 633, \dodoi{10.1051/0004-6361/200912013}

\bibitem[{{van Kempen} {et~al.}(2010){van Kempen}, {Kristensen}, {Herczeg}, {Visser}, {van Dishoeck}, {Wampfler}, {Bruderer}, {Benz}, {Doty}, {Brinch}, {Hogerheijde}, {J{\o}rgensen}, {Tafalla}, {Neufeld}, {Bachiller}, {Baudry}, {Benedettini}, {Bergin}, {Bjerkeli}, {Blake}, {Bontemps}, {Braine}, {Caselli}, {Cernicharo}, {Codella}, {Daniel}, {di Giorgio}, {Dominik}, {Encrenaz}, {Fich}, {Fuente}, {Giannini}, {Goicoechea}, {de Graauw}, {Helmich}, {Herpin}, {Jacq}, {Johnstone}, {Kaufman}, {Larsson}, {Lis}, {Liseau}, {Marseille}, {McCoey}, {Melnick}, {Nisini}, {Olberg}, {Parise}, {Pearson}, {Plume}, {Risacher}, {Santiago-Garc{\'\i}a}, {Saraceno}, {Shipman}, {van der Tak}, {Wyrowski}, {Y{\i}ld{\i}z}, {Ciechanowicz}, {Dubbeldam}, {Glenz}, {Huisman}, {Lin}, {Morris}, {Murphy}, \& {Trappe}}]{vankempen2010}
{van Kempen}, T.~A., {Kristensen}, L.~E., {Herczeg}, G.~J., {et~al.} 2010, \aap, 518, L121, \dodoi{10.1051/0004-6361/201014615}

\bibitem[{{Velusamy} {et~al.}(2007){Velusamy}, {Langer}, \& {Marsh}}]{velusamy2007}
{Velusamy}, T., {Langer}, W.~D., \& {Marsh}, K.~A. 2007, \apjl, 668, L159, \dodoi{10.1086/522929}

\bibitem[{{Watson} {et~al.}(2016){Watson}, {Calvet}, {Fischer}, {Forrest}, {Manoj}, {Megeath}, {Melnick}, {Najita}, {Neufeld}, {Sheehan}, {Stutz}, \& {Tobin}}]{watson2016}
{Watson}, D.~M., {Calvet}, N.~P., {Fischer}, W.~J., {et~al.} 2016, \apj, 828, 52, \dodoi{10.3847/0004-637X/828/1/52}

\bibitem[{{Wright} {et~al.}(2023){Wright}, {Rieke}, {Glasse}, {Ressler}, {Garc{\'\i}a Mar{\'\i}n}, {Aguilar}, {Alberts}, {{\'A}lvarez-M{\'a}rquez}, {Argyriou}, {Banks}, {Baudoz}, {Boccaletti}, {Bouchet}, {Bouwman}, {Brandl}, {Breda}, {Bright}, {Cale}, {Colina}, {Cossou}, {Coulais}, {Cracraft}, {De Meester}, {Dicken}, {Engesser}, {Etxaluze}, {Fox}, {Friedman}, {Fu}, {Gasman}, {G{\'a}sp{\'a}r}, {Gastaud}, {Geers}, {Glauser}, {Gordon}, {Greene}, {Greve}, {Grundy}, {G{\"u}del}, {Guillard}, {Haderlein}, {Hashimoto}, {Henning}, {Hines}, {Holler}, {Detre}, {Jahromi}, {James}, {Jones}, {Justtanont}, {Kavanagh}, {Kendrew}, {Klaassen}, {Krause}, {Labiano}, {Lagage}, {Lambros}, {Larson}, {Law}, {Lee}, {Libralato}, {Lorenzo Alverez}, {Meixner}, {Morrison}, {Mueller}, {Murray}, {Mycroft}, {Myers}, {Nayak}, {Naylor}, {Nickson}, {Noriega-Crespo}, {{\"O}stlin}, {O'Sullivan}, {Ottens}, {Patapis}, {Penanen}, {Pietraszkiewicz}, {Ray}, {Regan}, {Roteliuk}, {Royer}, {Samara-Ratna}, {Samuelson}, {Sargent}, {Scheithauer},
  {Schneider}, {Schreiber}, {Shaughnessy}, {Sheehan}, {Shivaei}, {Sloan}, {Tamas}, {Teague}, {Temim}, {Tikkanen}, {Tustain}, {van Dishoeck}, {Vandenbussche}, {Weilert}, {Whitehouse}, \& {Wolff}}]{wright2023}
{Wright}, G.~S., {Rieke}, G.~H., {Glasse}, A., {et~al.} 2023, \pasp, 135, 048003, \dodoi{10.1088/1538-3873/acbe66}

\bibitem[{{Yang} {et~al.}(2022){Yang}, {Green}, {Pontoppidan}, {Bergner}, {Cleeves}, {Evans}, {Garrod}, {Jin}, {Kim}, {Kim}, {Lee}, {Sakai}, {Shingledecker}, {Shope}, {Tobin}, \& {van Dishoeck}}]{yang2022}
{Yang}, Y.-L., {Green}, J.~D., {Pontoppidan}, K.~M., {et~al.} 2022, \apjl, 941, L13, \dodoi{10.3847/2041-8213/aca289}

\bibitem[{{Zhang} {et~al.}(2016){Zhang}, {Arce}, {Mardones}, {Cabrit}, {Dunham}, {Garay}, {Noriega-Crespo}, {Offner}, {Raga}, \& {Corder}}]{zhang2016}
{Zhang}, Y., {Arce}, H.~G., {Mardones}, D., {et~al.} 2016, \apj, 832, 158, \dodoi{10.3847/0004-637X/832/2/158}

\bibitem[{{Zhang} {et~al.}(2019){Zhang}, {Arce}, {Mardones}, {Cabrit}, {Dunham}, {Garay}, {Noriega-Crespo}, {Offner}, {Raga}, \& {Corder}}]{zhang2019}
---. 2019, \apj, 883, 1, \dodoi{10.3847/1538-4357/ab3850}

\end{thebibliography}
